\documentclass[12pt]{article}
\usepackage{mdframed}
\usepackage{float}
\usepackage{natbib}
\usepackage{multirow}
\usepackage{multicol}
 \usepackage{graphicx}
 \usepackage{mathtools}
 \usepackage{listings}
 \usepackage{bm}
 \usepackage{xcolor}
 \usepackage{setspace}
 \usepackage{pdfpages}
 \usepackage{amsmath,amsfonts}
 \usepackage{rotating}
  \usepackage[margin=1in]{geometry}
  \usepackage{subfigure}
  \usepackage{authblk}
  \usepackage{url}
\newcommand{\Ab}{\bm{A}}
\newcommand{\Db}{\bm{D}}
 \newcommand{\yb}{\bm{y}}

  \newcommand{\zb}{\bm{z}}

  \newcommand{\Bb}{\bm{B}}

  \newcommand{\Cb}{\bm{C}}  
  
  \newcommand{\Yb}{\bm{Y}}    
  \newcommand{\Ybt}{\widetilde{\bm{Y}}}  
  \newcommand{\Zb}{\bm{Z}}  
  \newcommand{\Zbt}{\widetilde{\bm{Z}}}
  \newcommand{\Kt}{\tilde{K}}
  \newcommand{\Wb}{\bm{W}}

  \newcommand{\gammab}{\bm{\gamma}}    
  \newcommand{\deltab}{\bm{\delta}}  
  \newcommand{\pib}{\bm{\pi}}
  \newcommand{\Lambdab}{\bm{\Lambda}}
  \newcommand{\Phib}{\bm{\Phi}} 
  \newcommand{\Eb}{\bm{E}}
  \newcommand{\Khat}{\widehat{K}}
  
    \newcommand{\alphab}{\bm{\alpha}}      
    \newcommand{\betab}{\bm{\beta}}  
    
    \newcommand{\thb}{\bm{\theta}}
    \newcommand{\longoverbrace}[2]{\overbrace{#1}^{\text{\hbox to 0cm{\hss #2 \hss}}}}  
    \newcommand{\longunderbrace}[2]{\underbrace{#1}_{\text{\hbox to 0cm{\hss #2 \hss}}}}

    \newcommand{\Poi}{\mbox{Poi}}
    \newcommand{\Cat}{\mbox{Cat}}

\definecolor{DarkGreen}{rgb}{0.5,0.8,0.6}   
\definecolor{RGBblack}{rgb}{0.0,0.0,0.0}    
\newcommand{\bch}{\color{black}}

\newcommand{\ech}{\rm\color{black}}

\newcommand{\blind}{0}
\begin{document}

\def\spacingset#1{\renewcommand{\baselinestretch}%
	{#1}\small\normalsize} \spacingset{1}


\if0\blind
{
	\title{\bf Bayesian Double Feature Allocation for Phenotyping with Electronic Health Records}
	\author[1,2]{Yang Ni}
	\author[3]{Peter M\"{u}ller}
	\author[4,5]{Yuan Ji}
	
	\affil[1]{Department of Statistics, Texas A\&M University}
	\affil[2]{Department of Statistics and Data Sciences, The University of Texas at Austin}
	\affil[3]{Department of Mathematics, The University of Texas at Austin}
	\affil[4]{Department of Public Health Sciences, The University of Chicago}
	\affil[5]{Program of Computational Genomics \& Medicine, NorthShore University HealthSystem}
	
	\date{}
	
	\maketitle
} \fi

\if1\blind
{
	\bigskip
	\bigskip
	\bigskip
	\begin{center}
		{\LARGE\bf Bayesian Double Feature Allocation for Phenotyping with Electronic Health Records}
	\end{center}
	\medskip
} \fi

\bigskip
\begin{abstract}
	We propose a categorical matrix factorization method to infer latent diseases from electronic health records (EHR) data in an unsupervised manner. A latent disease is defined as an unknown biological aberration that causes a set of common symptoms for a group of patients. The proposed approach is based on a novel double feature allocation model which simultaneously allocates features to the rows and the columns of a categorical matrix. Using a Bayesian approach, available prior information on known diseases greatly improves identifiability and interpretability of latent diseases. This includes known diagnoses for patients and known association of diseases with symptoms. We validate the proposed approach by simulation studies including mis-specified models and comparison with sparse latent factor models. In  the application to Chinese EHR data, we find interesting results, some of which agree with related clinical and medical knowledge. 
\end{abstract}

\noindent%
\vfill
\newpage
\spacingset{1.45} 

\section{Introduction}
\label{sec:intro}
Electronic health records (EHR) data electronically document medical diagnoses and clinical symptoms by the health care providers. The
digital nature of EHR automates access to health information and
allows physicians and researchers to take advantage of
a wealth
of data. Since its emergence, EHR has motivated novel data-driven approaches for
a wide range of tasks including phenotyping, drug assessment, natural
language processing, data integration, clinical decision support,
privacy protection and data mining \citep{ross2014big}. In this
article, we propose a double feature allocation (DFA) model for latent disease phenotyping.  A latent disease is defined as an
unknown biological aberration that causes a set of common symptoms for a group of
patients. The DFA model is a probability model on a categorical matrix
with each entry representing a symptom being recorded for a
specific patient.
The proposed model is based on an extension of
the Indian buffet process (IBP,
\citealt{ghahramani2006infinite}).
The generalization 
allows many-to-many
patient-disease and symptom-disease relationships and does not require
fixing the number of diseases \textit{a priori}.  Existing diagnostic
information is incorporated in the model to help identify and interpret latent
diseases. DFA can be also viewed as an alternative
representation of  categorical matrix
factorization  or as inference for an  edge-labeled random network. While recent phenotyping methods \citep{halpern2016electronic,henderson2017granite} are mostly performed via supervised learning, the proposed DFA is an unsupervised approach that aims to identify latent diseases.

The proposed approach builds on models for Bayesian inference of
hidden structure, including nonparametric mixtures, as reviewed, for example, in
\cite{favaro&teh:2013,barrios&al:2013}, 
graphical models 
\citep{green2011sampling,dobra2011bayesian},
matrix factorization 
\citep{rukat2017bayesian},
and random partitions and feature allocation  as discussed and
reviewed, for example, in 
\cite{broderick2013cluster} or \cite{campbell2018exchangeable}.
We explain below how graphs and matrix factorization relate to the
proposed model and inference. The proposed inference is motivated by EHR data that 
were collected  in routine  physical examinations for residents
in a city in  northeast  China in 2016.
The dataset contains both, information on some
diagnoses that were recorded  by the physicians,  as well as
symptoms from laboratory test results, such as metabolic and lipid panels.
The  diagnoses  are
binary variables indicating whether a patient has a disease or
not, and includes only some selected diseases.
Data on symptoms  are categorical variables with the number of
categories depending on the specific type of the symptom. For example,
heart rate is divided into three categories, low, normal and high,
whereas low density lipoprotein is classified as normal vs abnormal
(elevated) levels. The availability of both, diagnostic and symptomatic
information provides a good opportunity to detect latent diseases via
statistical modeling, that is, to infer latent disease information in
addition to the disease diagnoses that are already included in the data.
The proposed DFA  model  simultaneously allocates
patients and symptoms into the same set of latent features that are
interpreted as latent diseases.

Many  generic methods have been
developed for identifying latent patterns, which can be potentially
adopted for disease mining. We briefly review related
literature. 
\textit{Graphical models} \citep{lauritzen1996graphical} succinctly
describe a set of coherent conditional independence relationships of
random variables by representing their distribution as a
graph. Conditional independence structure can be directly read off
from the graph through the notion of graph separation. Graphical
models can reveal certain latent patterns of symptoms. For
example, by  extracting cliques (maximal fully connected subgraphs) from
an estimated graph of symptoms one can identify symptoms that are
tightly associated with each other. An underlying disease that is linked to 
these symptoms may explain the association. However, this
approach provides no {\it patient-level inference} since the patient-disease
relationship is not explicitly modeled and the choice of using cliques
rather than other graph summaries  remains
arbitrary.

\textit{Clustering models}  partition entities into
mutually exclusive latent groups (clusters).   Numerous methods have been
developed, including algorithm-based approaches such as k-means, and
model-based clustering methods such as finite mixture models
\citep{richardson1997bayesian,Miller2017} and infinite mixture models
\citep[for example]{lau2007bayesian,favaro&teh:2013,barrios&al:2013}.
%
Partitioning symptoms, similar to graphical models, may discover latent
diseases that are related to subsets of symptoms whereas clustering
patients suggests latent diseases that are shared among groups
of patients. Jointly clustering both symptoms and patients, also known
as bi-clustering
\citep{hartigan1972direct,
  li2005general,
  xu2013nonparametric}, allows one to simultaneously learn
patient-disease relationships and symptom-disease relationships.
The main limitation  of clustering methods
is the stringent assumption that each patient and symptom is related
to exactly one disease. Moreover, most biclustering methods deal with
continuous data \citep{guo2013bayesian}. 

\textit{Feature allocation models}
\citep{ghahramani2006infinite,broderick2013cluster}, also known as
overlapping clustering, relaxes the restriction to mutually exclusive subsets  
and allocates each unit to possibly more than one feature.
When simultaneously applied to both,
patients and symptoms, it is referred to as overlapping biclustering or
DFA.  Like biclustering, most of the existing DFA approaches only
handle continuous outcomes. See
\cite{pontes2015biclustering} for a recent review and references therein.
Only few methods, such as
\cite{wood2006npb} and \cite{uitert2008biclustering} can be applied to
discrete data,  but  are constrained to binary observations,
which is unsuitable for our application with categorical
observations. The proposed DFA extends existing methods  to  general
categorical data, automatic selection of the number of features, and
incorporation of prior information. 

\textit{Latent factor models} construct a low-rank representation of the
covariance matrix of a multivariate normal distribution. Latent factor
models assume that the 
variables are continuous and follow independent normal distributions
centered on latent 
factors multiplied by factor loadings. Imposing sparsity constraints
\citep{bhattacharya2011sparse,rovckova2016fast}, latent factor models
can be potentially adopted for the discovery of latent causes. However, the assumptions of normality and linear structure are often
violated in practice and it is not 
straightforward to incorporate known diagnostic information into
latent factor models. Principal component analysis has the same limitations.

\textit{Matrix factorization} is closely related to factor
models but not necessarily constrained to multivariate normal sampling
distribution. The most relevant variation of matrix factorization
for our application is binary/boolean matrix factorization
\citep{meeds2007modeling,zhang2007binary,miettinen2008discrete} which
decomposes a binary matrix into two low-rank latent binary
matrices. The proposed DFA can be viewed as categorical matrix
factorization which  includes  binary matrix factorization as a
special case. 

The contributions of this paper are three-fold: (1) we introduce a
novel categorical matrix factorization model based on DFA;
(2) we 
incorporate prior information to identify and interpret latent
diseases; and (3) we make inference on the number of latent diseases,
patient-disease relationships and symptom-disease relationships.

\subsection{Motivating case study: electronic health records}
\label{sec:mcs}
EHR data provides great opportunities for data-driven approaches in
 early disease detection,  screening and prevention. We consider EHR
data for $n=1000$ adults aged from 45 to 102 years with median 71
years. The dataset is based on physical examinations of residents in
some districts of a city in northeast China,
and was collected in 2016. The sample size of n=1000 corresponds to several weeks' worth of
data. In this paper we focus on developing model-based inference, including
full posterior simulation and summaries and therefore focus on a
moderate sample size. We will show how meaningful inference about
disease discovery is possible with such data. \bch Extension to larger datasets is discussed in Section \ref{sec:disc}. \ech
As in any work with EHR data, model-based inference needs to be
followed up by expert judgment to confirm the proposed latent diseases
and other inference summaries. Once confirmed, inferred disease
relationships become known prior information for future weeks.
This is how we envision an on-site implementation of the proposed methods.

\begin{figure} \centering
  \subfigure[]
  {\includegraphics[width=0.49\textwidth]{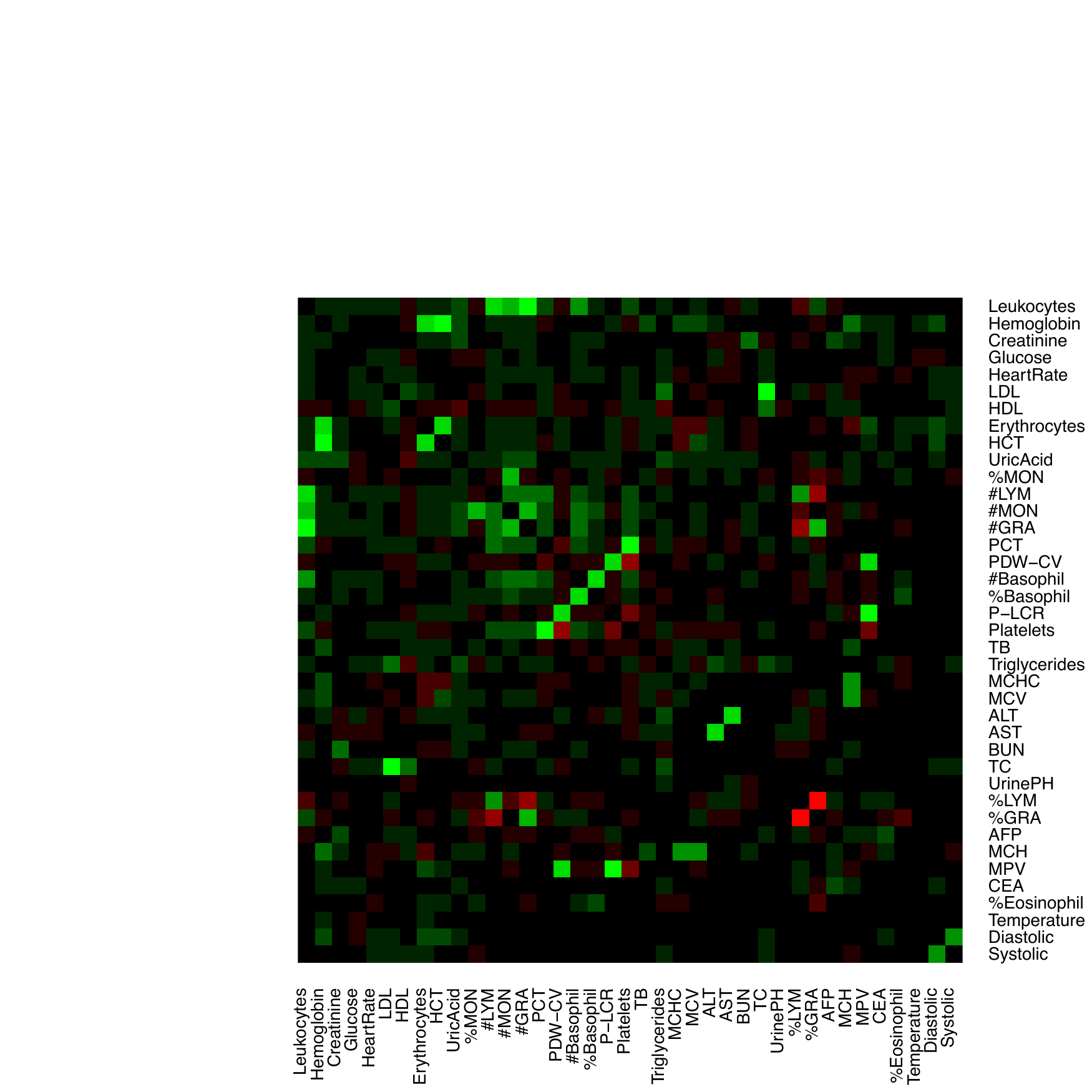}\label{fig:hmcov}}
  \subfigure[]
  {\includegraphics[width=0.49\textwidth]{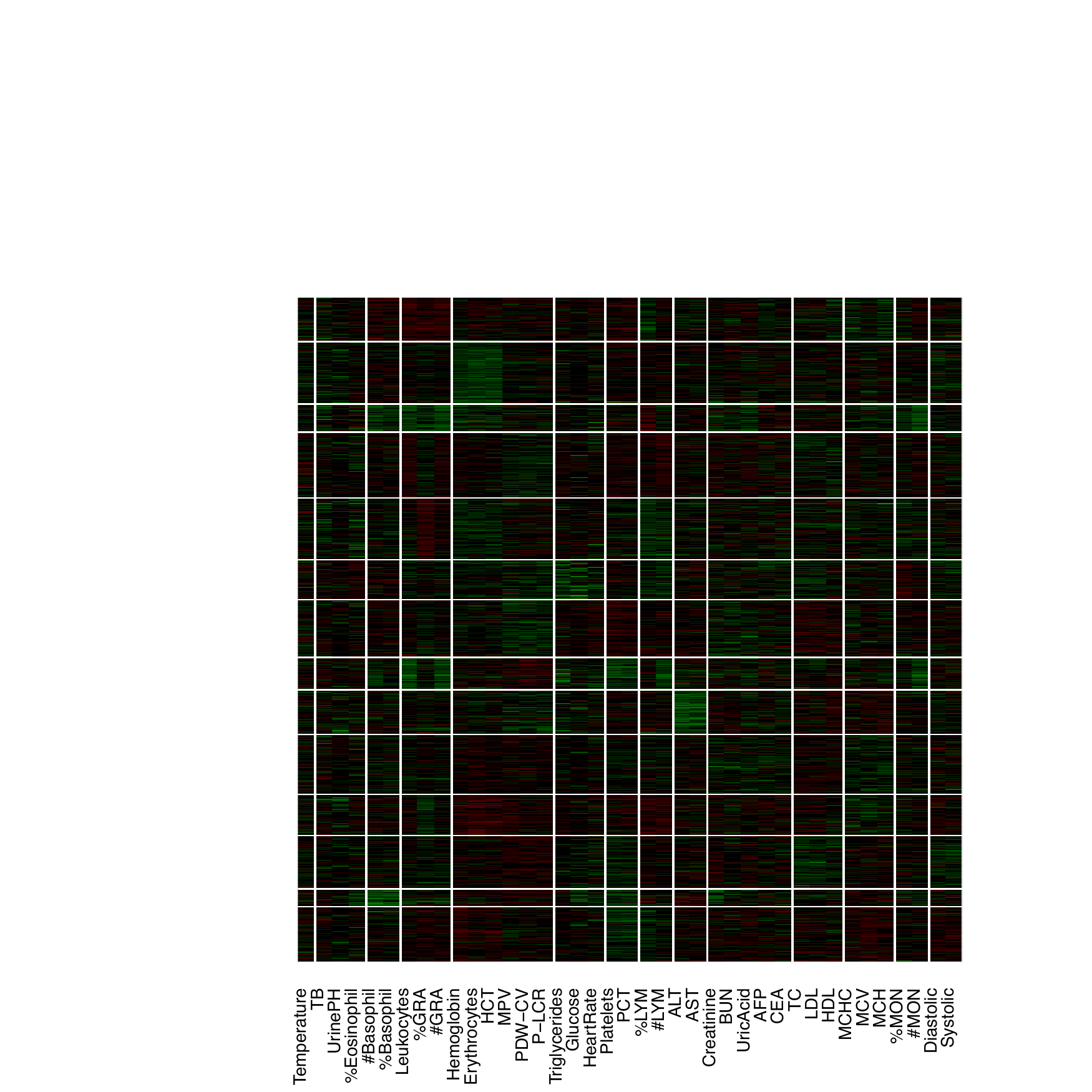}\label{fig:hmkm}}  
  \subfigure[]
  {\includegraphics[width=0.49\textwidth]{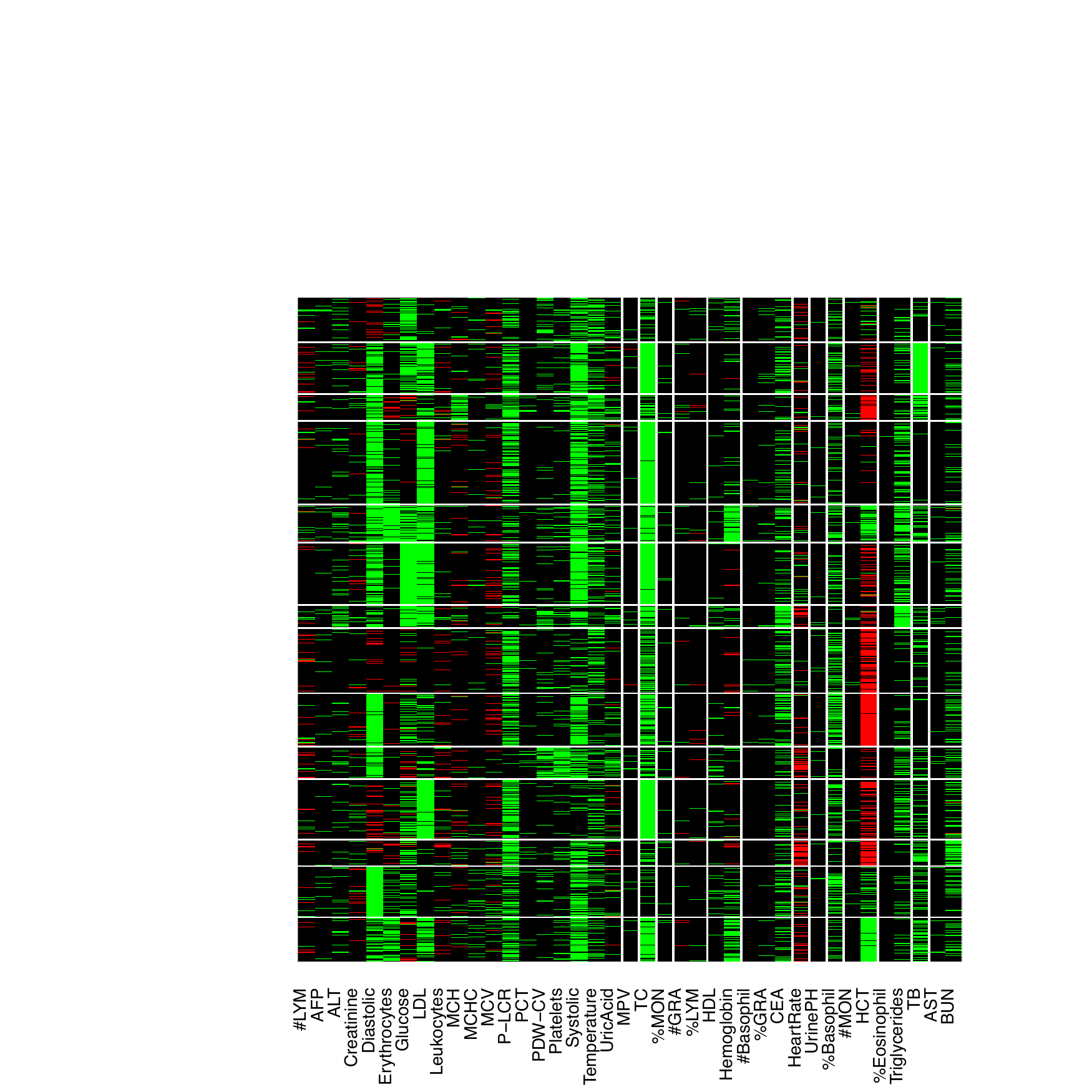}\label{fig:hmc}}
  \subfigure[]
  {\includegraphics[width=0.49\textwidth]{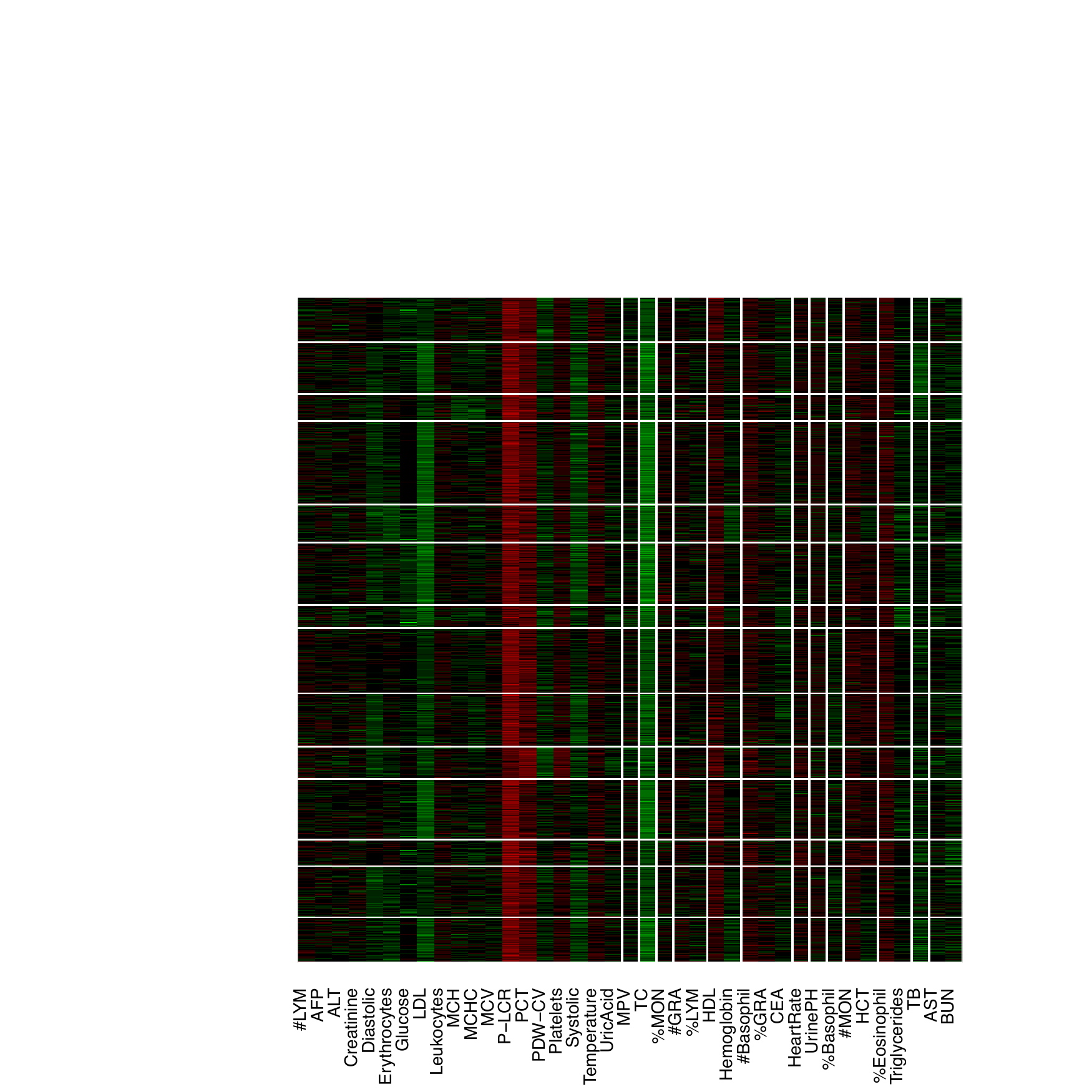}\label{fig:hmkm2}}
  \label{fig:hm}
  \caption{Heatmap of blood test data. In Panel (a), the
    correlation structure of the testing items is shown with green, black,
    and red colors indicating positive, negligible and negative
    correlations. Diagonal entries are set to 0. In Panels (b), (c) and (d),
    the rows and columns are ordered according to K-means with $K=14$. The columns are the testing items and the rows are
    patients. \bch In Panel (b), the EHR data is standardized. \ech Green, black and
    red colors represent values above, near and below the average or the midpoint. In Panel (c), the EHR data is transformed to a
    categorical matrix based on a trichotomization with respect
    to a reference range.  Green, black and red
    colors indicate above, within, below the range, respectively. \bch
    In panel (d), the EHR data is scaled and  centered at the midpoint
    of the reference range (the order of rows and columns
    is the same as in Panel (c)).  \ech
  }\label{fig:hms}
\end{figure}

The data contain blood test results measured on 39 testing
items which are listed in Table \ref{tab:names}.  
 Figure \ref{fig:hms}(a) shows 
the empirical correlation 
structure of the testing items as a heatmap with green,
black, and red colors indicating positive, negligible and negative
correlations. With appropriate ordering of the test items, 
 one can  see some patterns on the upper-left corner of the
heatmap. However, the patterns seem vague and
have no clear interpretation. The heatmap of the standardized data
is shown in Figure \ref{fig:hmkm} with green, black and red colors
indicating values above, near and below the average. Next we cluster the data using a
K-means algorithm
with $K=14$ (the number of latent diseases identified in later
 model-based inference), 
applied to both,  rows and columns  of the data matrix.
The clusters  find some interesting structures. 
For example,  indexing the submatrices in the heatmap by row and
column blocks, 
the values in block (9,9) tend to be
above  the average, whereas the values in the block (1,4) tend to be
below the average. However, there are at least two difficulties in
interpreting the clusters as latent diseases. Firstly, there is no
absolute relationship between the normal range of a testing item
and its population average. A deviation from the average does
not necessarily indicate an abnormality. Likewise,  average values of
testing items, especially those related to common diseases such as
hypertension,  are not necessarily within the normal range. For
instance, the mean and the median of systolic blood pressure in our
dataset are 147 mm Hg and 145 mm Hg, both of which are beyond the
threshold 140 mm Hg for hypertension (the high blood pressure values might be related to the elderly
patient population).
Secondly, the exploratory
analysis with K-means does not explicitly model patient-disease
relationships and symptom-disease relationships. For example,  one may be tempted
to interpret each column block as a latent disease. As a
consequence, each testing item has to be associated with exactly one
disease and the patient-disease relationship is unclear. If instead,
we define a latent disease by the row blocks, then each patient has to
have exactly one disease and the symptom-disease relationship is
ambiguous. 
	
We can slightly improve interpretability by incorporating prior
information. Specifically, each testing item comes with a reference
range which we use to define symptoms: a symptom is an item beyond the
reference range. In essence, we convert the original data matrix into
a ternary matrix which is shown in Figure \ref{fig:hmc}. The first
difficulty is resolved but the second difficulty remains. For
instance, the 6th column seems to suggest a disease with elevated
total cholesterol and low density lipoproteins, which is also found in our later
analysis with the proposed DFA. However, just as in Figure
\ref{fig:hmkm}, it is hard to judge which blocks meaningfully
represent latent disease since patient-disease relationships and
symptom-disease relationships are not explicitly modeled.
 Besides the requirement of 
specifying the number of clusters, K-means is unsuitable
for the task that we are pursuing in this paper. 

\bch Alternatively, instead of discretization, we can scale and center
test items at the midpoint of each reference range. We show the
heatmap in Figure \ref{fig:hmkm2} where the rows and the columns are
arranged in the same way as in Figure \ref{fig:hmc}. However, just
as previous cases, the same limitation of interpretability still
applies.  \ech

The proposed DFA addresses
all these issues, some of which have direct impact in
practice. For example, explicit modeling of patient-disease
relationships by DFA can be used to estimate the prevalence of latent
diseases in the target population, which is helpful for healthcare
policymakers. Finally, we emphasize 
that the gold standard of phenotyping
remains the judgment by trained clinicians. The inference from DFA,
however, is an important decision tool to facilitate this process, for instance, through the data-assisted personalized diagnosis support system (Section \ref{sec:shiny}).

	
The remainder of this paper is organized as follows. We introduce the
proposed DFA model in Section \ref{sec:dfam} and its alternative
interpretations in Section \ref{sec:ai}. Posterior inference is
described in Section \ref{sec:pi}.  Simulation studies and an EHR data
analysis are presented in Sections \ref{sec:ss} and \ref{sec:cs}. We
conclude this paper by a discussion in Section \ref{sec:disc}. 
	
\begin{table}	
  \caption{Blood test items. Acronyms are given within
    parentheses. Ternary symptoms (after applying the reference range)
    are in bold face. CV indicates coefficient of variation,
    ``dist'' is distribution, ``mn'' is mean,
    ``ct''is count, and ``conc" is concentration.} 
  \centering
	\scalebox{.8}{
\begin{tabular}{lll}
  \\\hline\hline
  alanine aminotransferase (ALT)& 
  aspartate aminotransferase (AST)&
  total bilirubin (TB)\\ 
  total cholesterol (TC)& 
  triglycerides& 
  low density lipoproteins (LDL)\\ 
  high density lipoproteins (HDL)& 
  urine pH (UrinePH)& 
  mn corpuscular hemoglobin conc (MCHC)\\
  \% of monocytes (\%MON) &
  alpha fetoprotein (AFP)& 
  carcinoembryonic antigen (CEA)\\ 
  number of monocytes (\#MON)&
  plateletcrit (PCT)& 
  CV of platelet dist. (PDW-CV) \\  
  \% of eosinophil (\%Eosinophil)&
  basophil ct (\#Basophil)& 
  \% of basophil (\%Basophil)\\
  platelet large cell ratio (P-LCR)& 
  platelets& 
  systolic blood pressure (Systolic)\\ 
  \% of granulocyte (\%GRA)& 
  body temperature (BodyTemperature)&
  \textbf{leukocytes}\\
  \textbf{hemoglobin}& 
  \textbf{creatinine}& 
  \textbf{blood urea nitrogen (BUN)}\\ 
  \textbf{glucose}& 
  \textbf{diastolic blood pressure (Diastolic)}& 
  \textbf{heart rate (HeartRate)}\\
  \textbf{erythrocytes}& 
  \textbf{hematocrit (HCT)}& 
  \textbf{uric acid (UricAcid)}\\ 
  \textbf{\% of lymphocyte (\%LYM)}&
  \textbf{mn corpuscular volume (MCV)}& 
  \bf mn corpuscular hemoglobin (MCH)\\ 
  \textbf{lymphocyte ct (\#LYM)}& 
  \textbf{granulocyte ct (\#GRA)}&  
  \textbf{mn platelet volume (MPV)} 
\end{tabular}}
	\label{tab:names}
\end{table}

\section{Double Feature Allocation Model}
\label{sec:dfam}
DFA can be applied to any 
categorical matrix. 
 For simplicity, and in anticipation of the application to EHR, 
we will describe our model for binary and ternary
matrices.
Let $y_{il}\in \{-1,0,+1\}$  and $z_{ij}\in \{0,1\}$ for
$i=1,\dots,n$, $l=1,\dots,q$, and $j=1,\dots,p$.  In the EHR dataset,
$y_{il}$ denote the observation for patient $i$
 for a symptom that  can be naturally
trichotomized into low ($y_{il}=-1$), normal ($y_{il}=0$) and high
($y_{il}=+1$) levels.  Similarly,   $z_{ij}$ denotes a symptom that is
dichotomized  into normal ($z_{ij}=0$) vs abnormal ($z_{ij}=1$)
levels. We assume that a patient experiences certain symptoms when
he/she has diseases that are related to those symptoms. In other
words, the patient-symptom relationships are thought to be generated
by two models: a patient-disease model and a symptom-disease model. 

\subsection{Patient-disease (PD) model}
The patient-disease relationships are defined by a binary matrix
$\Ab=(\alpha_{ik})\in \{0,1\}^{n\times \tilde{K}}$ with
$\alpha_{ik}=1$ meaning patient $i$ has disease $k$.
 We start the model construction assuming 
a fixed number $\Kt$ of diseases, to be relaxed later (and
we reserve the notation $K$ for the relaxation). 
Conditional on
$\Kt$, $\alpha_{ik}$ are assumed to be independent Bernoulli
random variables, $\alpha_{ik} \mid \pi_k\sim Ber(\pi_k)$ with $\pi_k$
following a conjugate beta prior, $\pi_k\sim Beta(m/\Kt,1)$. Here
$m$ is a fixed hyperparameter. Marginalizing out $\pi_k$,
$$
   p(\Ab)= \prod_{k=1}^{\Kt}
   \frac{m\Gamma(r_k+\frac{m}{\Kt})\Gamma(n-r_k+1)}
        {\Kt\, \Gamma(n+1+\frac{m}{\Kt})},
$$
where $r_k=\sum_{i=1}^n\alpha_{ik}$ is the sum of the $i$th column of
$\Ab$.\\
\indent  However, the number of latent diseases is unknown in practice
and inference on $\Kt$ is of interest. Let $H_n=\sum_{i=1}^n1/i$ be the
$n$-th harmonic number.  Next, take the limit $\Kt\rightarrow \infty$ and
remove columns of $\Ab$ with all zeros. \bch This step is not trivial
and needs careful  consideration  of equivalence classes of
binary matrices; details can be found in Section 4.2 of
\cite{griffiths2011indian}. \ech  Let $K$ denote the number of 
non-empty columns.
The resulting matrix $\Ab$ follows an $\mbox{IBP}(m)$ prior
(without a specific column ordering), 
with probability
\begin{equation}
  \label{ibp}
  p(\Ab)=
  \frac{m^{K}\exp\{-m H_n\}} {K!}
  \prod_{k=1}^{K}\frac{\Gamma(r_k)\Gamma(n-r_k+1)}
                     {\Gamma(n+1)}.
\end{equation}
Since $K$ is  random and unbounded,  we do not need to specify
the number of latent diseases \textit{a priori}. And with a finite
sample size, the number $K$ of non-empty diseases is finite with
probability one.
\bch See, for example, \cite{griffiths2011indian} for a review of the
IBP, including the motivation for the name based on a generative
model. We only need two results that are implied by this generative
model.
Let $r_{-i,k}$ denote the sum in column $k$, excluding row $i$. \ech
Then the conditional probability  for $\alpha_{ik}=1$  is
\begin{align}
\label{eqn:pc}
p(\alpha_{ik}=1 \mid \alphab_{-i,k})=r_{-i,k}/n,
\end{align}
provided $r_{-i,k}>0$, where $\alphab_{-i,k}$ is the $k$th column of
$\Ab$ excluding $i$th row.
And the distribution of the number of new features
(disease) for each row (patient) is $\Poi(m/n)$.

\subsection{Symptom-disease (SD) model}
A disease may trigger multiple symptoms and a symptom may be  related to multiple diseases.
The SD model allows both. Let  $K$ again denote the number of latent diseases. The SD model inherits $K$ from the PD model. For binary symptoms, we generate a binary  $(p \times K)$  matrix
 $\Bb=[\beta_{jk}]$, $\beta_{jk}\in \{0,1\}$,  from independent Bernoulli
distributions, $\beta_{jk}\sim Ber(\rho)$ with $\rho\sim
Beta(a_\rho,b_\rho)$.  Similarly, for ternary symptoms, we generate a
 $(q \times K)$  ternary matrix  $\Cb=[\gamma_{lk}]$,
$\gamma_{lk}\in\{-1,0,1\}$, from 
independent categorical distributions $\gamma_{lk}\sim Categ(\pib)$ 
with $\pib\sim Dir(\phi_{-1},\phi_0,\phi_{+1})$.  

\bch We have considered that $\beta_{jk}\sim Ber(\rho)$ and $\rho\sim
Beta(a_\rho,b_\rho)$, that is, all binary symptoms (similarly for
ternary symptoms) have the same probability of being triggered by any
disease \textit{a priori}, which implies
that the $\beta_{jk}$'s are exchangeable across symptoms and diseases.  The construction with the unknown hyperparameter $\rho$ 
allows for automatic multiplicity control 
\citep{scott2010bayes}. However,  exchangeability does not necessarily
hold in all cases. For example, symptoms such as fever are more common
than symptoms such as chest pain. If desired, such heterogeneity
across symptoms can be incorporated in the proposed model by setting
$\beta_{jk}\sim Ber(\rho_j)$. In addition, the beta prior on $\rho_j$
can be adjusted according to the prior knowledge of the importance of
each symptom $j$.
 Letting $\lambda_j = \Phi^{-1}(\rho_j)$ denote a probit
transformation of $\rho_j$, one can introduce dependence across
symptoms by a hierarchical model, 
\begin{gather}
\label{eqn:hprior}
  \lambda_j \mid \lambda, \sigma^2  \sim  N(\lambda,\sigma^2)\\
  \lambda  \sim  p(\lambda)\propto1,~ \sigma^2\sim
                   IG(a_\sigma,b_\sigma)
 \nonumber                   
\end{gather}
We explore this approach  in a simulation study
reported in the Supplementary Material A.
A similar strategy can be used to accomodate 
heterogeneity across diseases, in which case we assume
$\beta_{jk}\sim Ber(\rho_k)$. 
\ech

Note that $\Ab$, $\Bb$ and $\Cb$ have the same number of columns
because they share the same latent diseases.  In the language of the
Indian restaurant metaphor, each dish (disease) corresponds to a
combination of ingredients (symptoms). Some ingredients come at spice
levels $\{-1,1\}$, if selected.
 We have now augmented model \eqref{ibp} by matching each subset of
patients, i.e., each disease, with a subset of symptoms defined in 
$\Bb$ and $\Cb$. As a result, each disease, or feature, is defined as a
pair of random subsets of patients and symptoms, respectively.
We therefore refer to the model as double feature allocation (DFA).
\bch In this construction we have defined the joint prior of $\Ab,\Bb$ and
$\Cb$ using the factorization $p(\Ab,\Bb,\Cb)=p(\Ab)p(\Bb,\Cb \mid \Ab)$,
i.e. first assuming an IBP prior for $\Ab$, and then defining a prior on
$\Bb$ and $\Cb$ conditional on $\Ab$ (through $K$, the number of
columns of $\Ab$). In principle, the joint prior can
 alternatively be factored as 
$p(\Ab,\Bb,\Cb)=p(\Bb,\Cb)p(\Ab \mid \Bb,\Cb)$, in which case
we start the construction with a prior $p(\Bb,\Cb)$. 
However, the categorical nature of $\Cb$ complicates this construction
and we therefore prefer the earlier factorization. 

 The main variations in the construction, compared to traditional
use of the IBP as prior for a binary matrix, is the use of matched
subsets of patients and symptoms for each feature, the mix of binary
and categorical items, and the specific sampling model, as it arises
in the motivating application. \ech

\subsection{Sampling model}
Once we generate the PD and SD relationships, the observed data matrix
which records  
the symptoms for each patient is generated by the following sampling models. Let
$\alphab_i$ denote the $i$th row of $\Ab$, $\betab_j$ the $j$th row of
$\Bb$ as a column vector, and $\gammab_l$ the $l$th row of $\Cb$, written  as a column vector. We assume conditionally
independent Bernoulli distributions for $z_{ij}$ 
\begin{align}
\label{eqn:zij}
&z_{ij}|\alphab_i, \betab_j, \Wb_j, \zeta_j \sim Ber\left\{\frac{\exp(\alphab_i\Wb_j\betab_j+\zeta_j)}{1+\exp(\alphab_i\Wb_j\betab_j+\zeta_j)}\right\},
\end{align}
with $\Wb_j=diag(w_{j1},\dots,w_{jK})$ where $w_{jk}$ is constrained
to be positive, so that the probability of
experiencing symptom $j$ for patient $i$ always increases
if a patient has a disease $k$ that triggers the symptom. 
The parameter $\zeta_j$ captures the remaining probability of symptom
$j$ that is unrelated to any disease.
 One could alternatively include the weight already in $\Bb$ (and
$\Cb$). We prefer separating the formation of the random subsets which
define the features versus  the weights which appear in the sampling
model. 

Similarly,  we assume conditionally independent categorical distributions
for $y_{il}$.
 Let $\Cat(\pi_1,\pi_2,\pi_3)$ denote a categorical distribution
 with probabilities $\pi_1,\pi_2,\pi_3$ for three outcomes.
Also let $\gammab_l^+=I(\gammab_l=+1)$ and $\gammab_l^-=I(\gammab_l=-1)$ with
$I(\cdot)$ being the element-wise indicator function. 
We assume
\begin{equation}
  y_{il} \mid \alphab_i, \gammab_l,  \Wb_l^-,\Wb_l^+, \eta_l^+,\eta_l^- \sim
  \Cat\left(Me^{\alphab_i\Wb_l^- \gammab_l^-+\eta_l^-},\; M, \;
  M e^{\alphab_i\Wb_l^+ \gammab_l^++\eta_l^+}\right),
\label{eqn:cat}
\end{equation}
with $M$ being the normalization constant,
$\Wb_l^-=diag(w_{l1}^-,\dots,w_{lK}^-)$ and
$\Wb_l^+=diag(w_{l1}^+,\dots,w_{lK}^+)$, where $w_{lk}^-,w_{lk}^+$ are
also constrained to be positive,  and $\eta_l^-$ and $\eta_l^+$ have
interpretations similar to $\zeta_j$.

We complete the model by assigning  vague  priors on
hyperparameters, $\zeta_j$, $\eta_l^-$, $\eta_l^+\sim N(0,\tau^2)$
with $\tau=100$ and $w_{jk}$, $w_{lk}^-$, $w_{lk}^+\sim
Gamma(1,\tau_w)$ with variance $\tau_w^2=100$. \bch A brief sensitivity
analysis for the choice of these hyperparameters is shown  in
Supplementary Material A. \ech

\subsection{Incorporating prior knowledge}
\label{sec:24}
Available diagnostic information is easily incorporated in the proposed
model.
 We fix the first $K_0$ columns of
$\Ab$ to represent available diagnoses related to $K_0$ known
diseases. \bch Specifically, we 
label the known diseases as $1,\dots,K_0$ and set $\alpha_{ik}=1$ if
and only if individual $i$ is diagnosed with disease $k$ for
$k=1,\dots,K_0$. \ech 
Similarly, known SD relationships are represented 
by fixing corresponding  columns of $\Bb$ or $\Cb$.  

\begin{figure}[h]
	\centering
	\includegraphics[width=.3\textwidth]{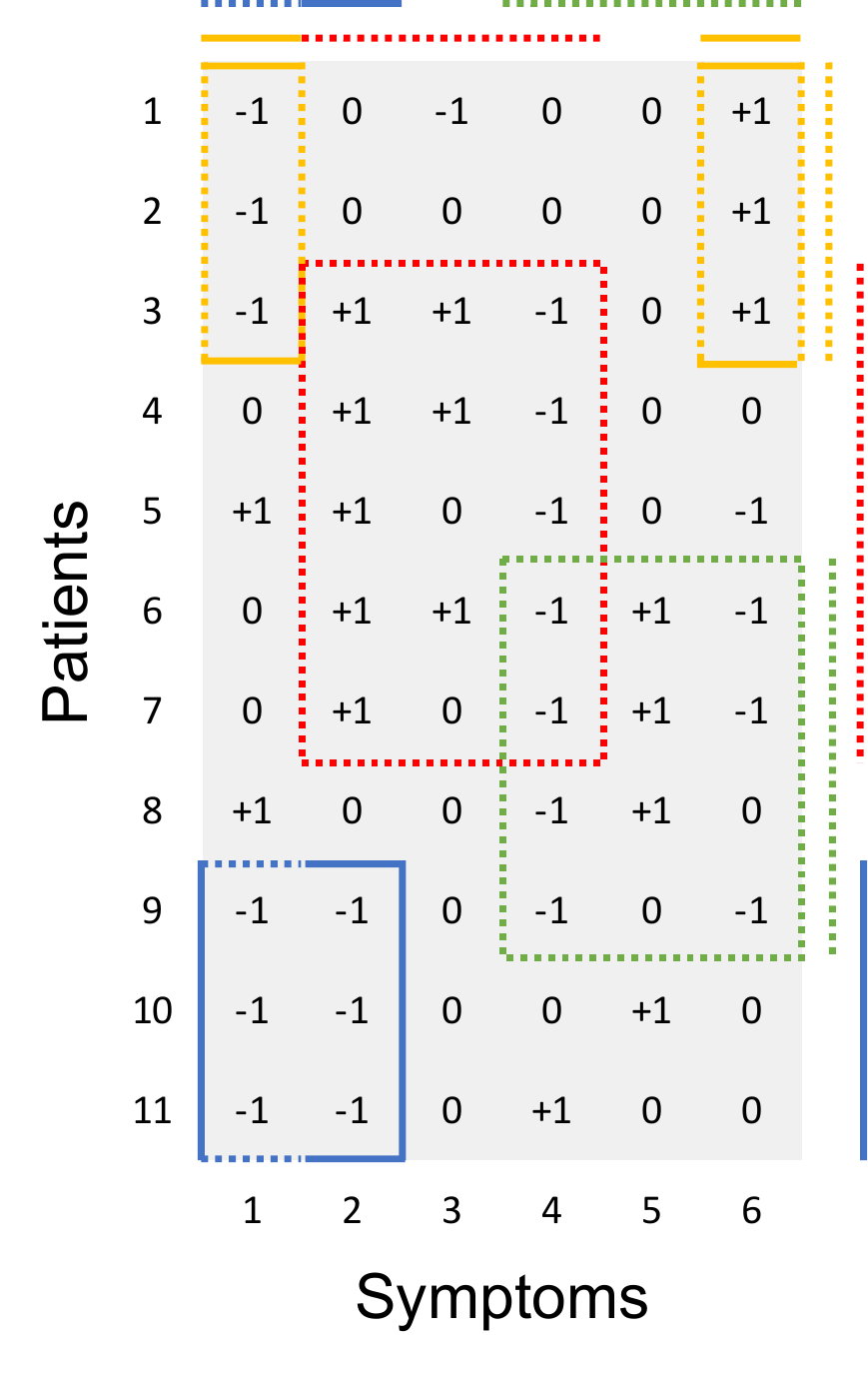}
	\caption{Illustration of DFA for ternary symptoms. The lines
          on the side/top indicate the grouping of patients/symptoms by
          diseases. Each disease is represented by one color. Dashed
          lines are latent whereas solid lines are known.} 
	\label{fig:dfa}
\end{figure}

A simple example with 11 patients and 6 ternary symptoms is shown in Figure
\ref{fig:dfa} for illustration. There are 4 diseases, each represented
by one color (corresponding to the dashed blocks inside the
matrix in Figure \ref{fig:dfa}).  Importantly, patients and symptoms can be linked to multiple
diseases. For example, patient 9 has both, the blue and the green
disease, and
symptom 4 can be triggered by either the red or the green disease.
Available prior information is incorporated in this example. For
instance, if patients 9, 10, 11 are diagnosed with the blue disease,
they will be grouped together deterministically (represented by the
blue solid line on the side). Likewise, if the yellow disease is known to lead to symptoms 1 and 6, we fix them in the model
(represented by the yellow solid lines on the top). 

\section{Alternative Interpretations}
\label{sec:ai}
The proposed DFA is closely related to matrix factorization and random
networks.
 We briefly discuss 
two alternative interpretations of DFA  for the case of
the observed data being ternary.  Binary outcomes are a special case of
ternary outcomes; generalization to more than three categories is
straightforward. We use the same toy example as in Section \ref{sec:24}
to illustrate the alternative interpretations. 

\begin{figure}
	\centering
	\includegraphics[width=.8\textwidth]{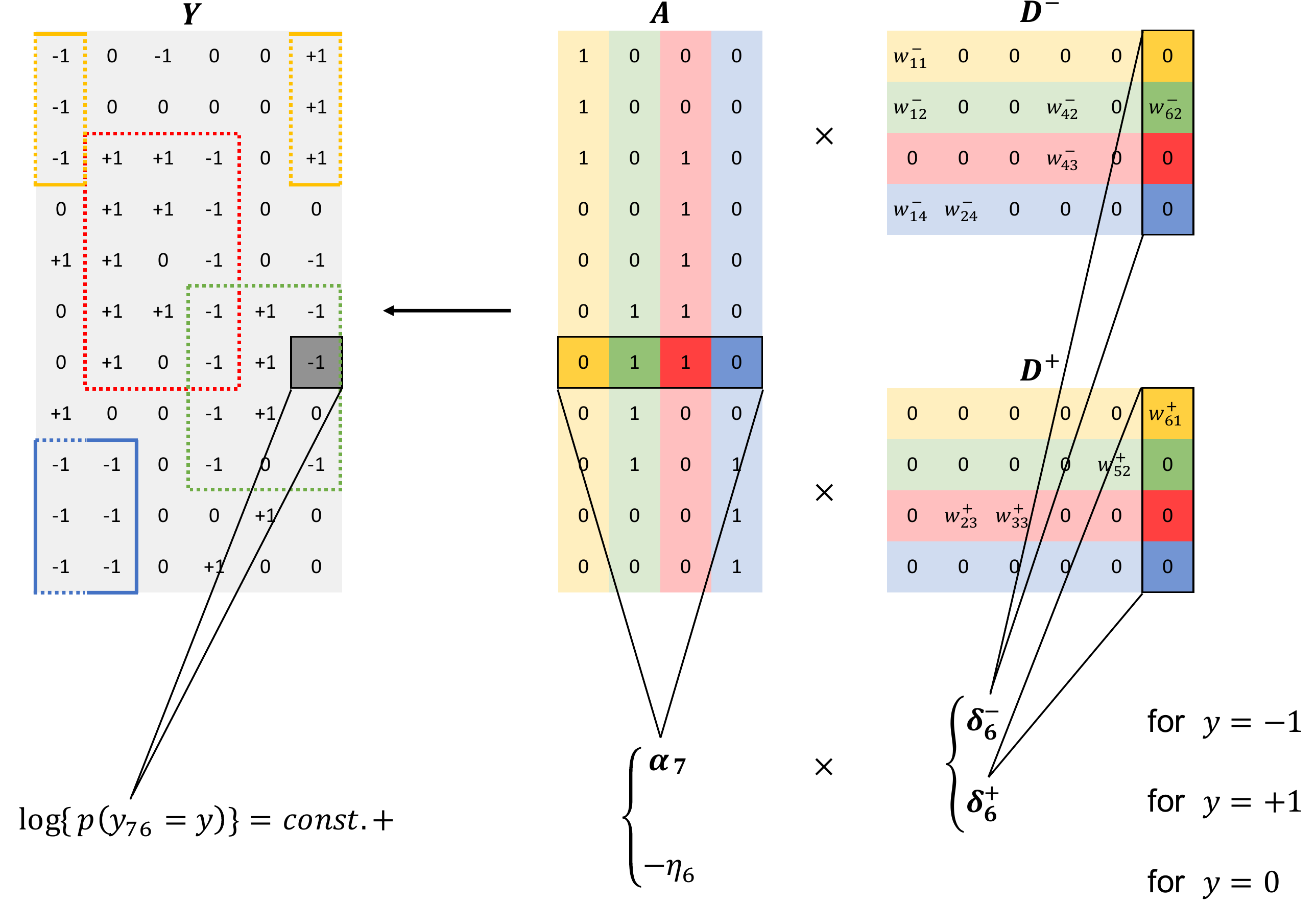}
	\caption{Categorical matrix factorization as an alternative representation of DFA in Figure \ref{fig:dfa}. $\Db^-=\Db\circ I(\Cb=-1)$ and $\Db^+=\Db\circ I(\Cb=+1)$. Diseases correspond to the columns of $\Ab$ and the rows of $\Db^-$ and $\Db^+$. They are represented by the same set of colors as in Figure \ref{fig:dfa}.}
	\label{fig:cmf}
\end{figure}

\textit{Categorical matrix factorization (CMF).} DFA can be viewed as
a CMF. Merging $\Wb$ and $\gammab$, model (\ref{eqn:cat})
probabilistically factorizes an $(n\times q)$ categorical matrix $\Yb$
into an $(n\times K)$ low-rank binary matrix $\Ab$ and an $K\times q$
low-rank nonnegative matrix $\Db=(\delta_{kl})$ where $K<\min(n,q)$,
$\delta_{kl}=w_{lk}^-\gamma_{lk}^-+w_{lk}^+\gamma_{lk}^+$,
$\gamma_{lk}^-=I(\gamma_{lk}=-1)$ and
$\gamma_{lk}^+=I(\gamma_{lk}=+1)$. From (\ref{eqn:cat}),  
\begin{eqnarray}
  \log\{p(y_{il}=y)\}=c+
  \left\{ \begin{array}{lcl}\alphab_i\deltab_l^-+\eta_l^-&\mbox{for}&y=-1\\\alphab_i\deltab_l^++\eta_l^+&\mbox{for}&y=+1\\0&\mbox{for}&y=0\end{array} \right.
\end{eqnarray}
where $\deltab_l=(\delta_{1l},\dots,\delta_{Kl})^T$,
$\deltab_l^-=\deltab_l\circ I(\gammab_l=-1)$,
$\deltab_l^+=\deltab_l\circ I(\gammab_l=+1)$ with element-wise
multiplication $\circ$, and
$c=-\log\{\exp(\alphab_i\deltab_l^-+\eta_l^-)+\exp(\alphab_i\deltab_l^++\eta_l^+)+1\}$. Figure
\ref{fig:cmf} illustrates the factorization. The matrix
$\Ab$ describes the PD
relationships. $\Db^-$ and $\Db^+$ characterize the SD relationships
where $\Db^-=\Db\circ I(\Cb=-1)$ and $\Db^+=\Db\circ I(\Cb=+1)$. The number of diseases is less than the number of patients
and the number of symptoms. 

\begin{figure}
	\centering
	\includegraphics[width=.7\textwidth]{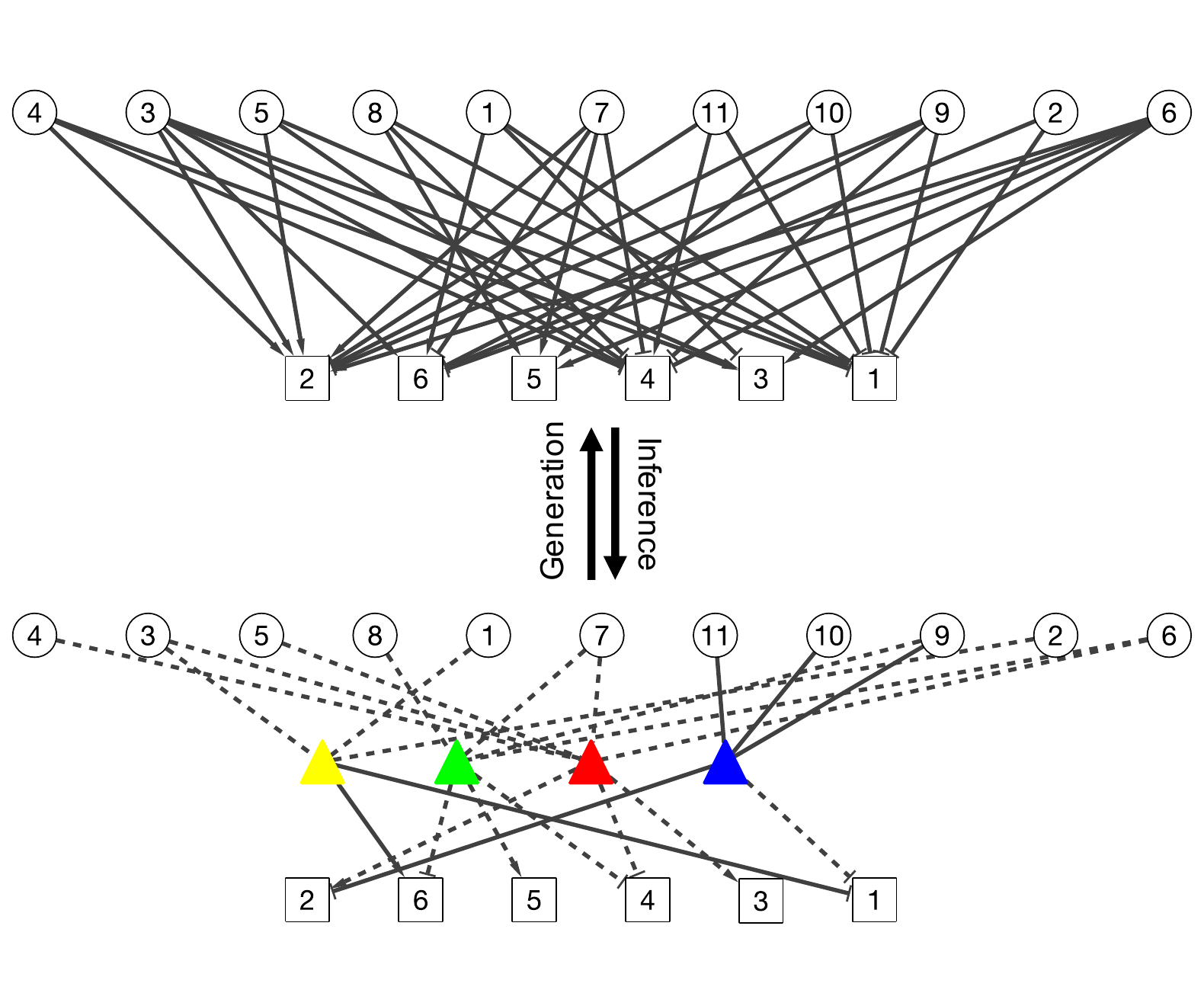}
	\caption{Edge-labeled random networks as an alternative representation of the DFA in Figure \ref{fig:dfa}. The observed bipartite graph (top) is assumed to be generated by the latent tripartite graph (bottom). DFA addresses the inverse problem. Circles are patients, squares are symptoms and triangles are diseases whose colors have the same interpretations as in Figure \ref{fig:dfa}. Dashed lines are latent whereas solid lines are known/observed. An undirected edge connecting patients and diseases is binary. Edges with arrow head or flat bars represent different types of SD relationships. }
	\label{fig:net}
\end{figure}

\textit{Edge-labeled random networks.} DFA can be also interpreted as
inference for a random network with labeled edges. The 
observed categorical matrix 
$\Yb$ is treated as a categorical adjacency matrix which encodes a
bipartite random network. Patients form one set of nodes and symptoms
form another set. The edge labels correspond to the categories in
$\Yb$. See the bipartite network on the upper portion of Figure
\ref{fig:net} where the two labels ($+1$, $-1$) are respectively represented by arrow heads and flat bars. DFA assumes that the observed bipartite
graph is generated from a latent tripartite graph (given in the lower
portion of Figure \ref{fig:net}).  Inference under the DFA
model reverses  the data generation process. The tripartite graph
 introduces an additional set of (latent) 
nodes corresponding to diseases. The edges
between patients (symptom) and diseases indicate PD (SD)
relationships. Prior PD and SD knowledge is  represented  by
fixing the corresponding edges (solid lines in Figure
\ref{fig:net}). 

\section{Posterior Inference}
\label{sec:pi}
The model described in Section \ref{sec:dfam} is parameterized by
$$
\thb=\{\Ab,\Bb,\Cb,\{\Wb_j,\zeta_j\}_{j=1}^p,\{\Wb^-_l,\Wb^+_l,\eta_l^-,\eta_l^+\}_{l=1}^q\}.
$$
Posterior
inference is carried out by
Markov chain Monte Carlo 
(MCMC) posterior simulation.  All parameters except $\Ab$ can be
updated with simple 
Metropolis-Hasting transition probabilities. Sampling $\Ab$ is
slightly more complicated because the dimension of the parameter space
can change. We therefore provide details  of the transition
probability for updating $\Ab$. \bch In the implementation,
for easier bookkeeping,  we set a large upper bound
$K_{\vee}=50$ for the number of latent diseases;
it is never reached during the course of the MCMC.
\ech
\paragraph*{MCMC} ~\\
 We use superscript $^{(t)}$ to index the state $\thb$ across
iterations. 
Initialize $\thb^{(0)}$. For $t=0,\dots,T-1$, do
\begin{itemize}
\item [(1)] Update $\Ab^{(t+1)}$. We scan through each row,
  $i=1,\dots,n$,  of $\Ab=\Ab^{(t)}$. 
  \begin{itemize}
  \item [(1a)] Update existing diseases $k=1,\dots,K$. If
    $\alphab_{-i,k}=\bm{0}$,  drop disease $k$; 
    otherwise, sample $\alpha_{ik}$ from the full conditional
    distribution. \bch For $j=0,1$ 
    \begin{multline}
      p(\alpha_{ik}=j\mid \cdot) \propto
      p(\alpha_{ik}=j \mid \alphab_{-i,k})\;
      p(\zb_i \mid \alpha_{ik}=j, \alphab_{i,-k},\Bb,\{\Wb_j,\zeta_j\}_{j=1}^p)\;
      \times\\
      p(\yb_i \mid \alpha_{ik}=j, \alphab_{i,-k},\Cb,\{\Wb^-_l,\Wb^+_l,\eta_l^-,\eta_l^+\}_{l=1}^q)\\ 
       = p(\alpha_{ik}=j \mid \alphab_{-i,k})
                                       \prod_{j=1}^pp(z_{ij} \mid
      \alpha_{ik}=j, \alphab_{i,-k},\betab_j,\Wb_j,\zeta_j)~ \times \\
      \prod_{l=1}^qp(y_{il} \mid
      \alpha_{ik}=j, \alphab_{i,-k}, \gammab_l,  \Wb_l^-,\Wb_l^+,
      \eta_l^-,\eta_l^+)
      \nonumber
    \end{multline}
    where the factors in the last equation are defined in
    \eqref{eqn:pc}, \eqref{eqn:zij} and \eqref{eqn:cat},
    respectively.\ech 
  \item [(1b)] Propose new diseases. The proposed new
    diseases are unique to patient $i$ only, 
    i.e., $\alpha_{ik}=1$ and $\alphab_{-i,k}=\bm{0}$. 
    We first draw $k^*\sim
    \mbox{Poi}(m/n)$. If $k^*=0$, go to the next step. Otherwise, we
    propose a set of new disease-specific parameters
    $\betab_k^*=(\beta_{1k}^*,\dots,\beta_{pk}^*)^T$,
    $\gammab_k^*=(\gamma_{1k}^*,\dots,\gamma_{qk}^*)^T$,
    $\{w_{jk}^*\}_{j=1}^p$, $\{w_{lk}^{-*},w_{lk}^{+*}\}_{l=1}^q$ for
     $k= K+1,\dots,K+k^*$  
     from their respective prior distributions. 
     We accept the new disease(s) and the disease-specific parameters, with
    probability 
    \begin{eqnarray*}
      \min\left\{1,
        \frac
        {\prod_{j=1}^pp(z_{ij} \mid \alphab_i^*,\betab_j^*,\Wb_j^*,\zeta_j)}
        {\prod_{j=1}^pp(z_{ij} \mid \alphab_i^{(t)},\betab_j^{(t)},\Wb_j^{(t)},\zeta_j)}\,
        \frac
        {\prod_{l=1}^qp(y_{il} \mid \alphab_i^*,\gammab_l^*,  \Wb_l^{-*},\Wb_l^{+*},
          \eta_l^-,\eta_l^+)}
        {\prod_{l=1}^qp(y_{il} \mid \alphab_i^{(t)}, \gammab_l^{(t)},  \Wb_l^{-(t)},\Wb_l^{+(t)},
          \eta_l^-,\eta_l^+)}\, \right\},
       \end{eqnarray*}
       
    where $\alphab_i^*=(\alphab_i^{(t)},1,\dots,1)$,
    $\betab_j^*=(\betab_j^{(t)T},\beta_{j,K+1}^*,\dots,\beta_{j,K+k^*}^*)^T$,
    $\gammab_l^*=(\gammab_l^{(t)T},\gamma_{l,K+1}^*,\dots,\gamma_{l,K+k^*}^*)^T$,
    $\Wb_j^*=blkdiag(\Wb_j^{(t)},w_{j,K+1}^*,\dots,w_{j,K+k^*}^*)$,
    $\Wb_l^{-*}=blkdiag(\Wb_l^{-(t)},w_{l,K+1}^{-*},\dots,w_{l,K+k^*}^{-*})$ and
    $\Wb_l^{+*}=blkdiag(\Wb_l^{+(t)},w_{l,K+1}^{+*},\dots,w_{l,K+k^*}^{+*})$. Note
    that the acceptance probability only involves the likelihood ratio
    because prior and proposal  probabilities cancel out. 
    If the new disease is accepted, we increase $K$ by $k^*$. 
  \end{itemize}
\item [(2)] Update all other parameters in $\thb^{(t+1)}$ using
  Metropolis-Hasting transition probabilities. 
\end{itemize}

To summarize the posterior distribution based on the  the
MCMC simulation output, we
proceed by first calculating the maximum a 
posteriori (MAP) estimate $\widehat{K}$ from the marginal posterior
distribution of $K$. Conditional on $\widehat{K}$, we find the least
squares estimator 
$\widehat{\Ab}$ by the following procedure \citep{dahl2006model,lee2015}.  For any two binary
matrices $\Ab,\Ab'\in \{0,1\}^{n\times \widehat{K}}$, we define a
distance
$d(\Ab,\Ab')=\min_{\pi}\mathcal{H}(\Ab,\pi(\Ab'))$ where
$\pi(\Ab')$ denotes a permutation of the columns of $\Ab'$ and
$\mathcal{H}(\cdot,\cdot)$ is the Hamming distance of two binary
matrices. A point estimate $\widehat{\Ab}$ is then obtained as 
\[
   \widehat{\Ab}=\arg\min_{\Ab'}\int d(\Ab,\Ab')dp(\Ab \mid
   \Zb,\Yb,\widehat{K}).
\]
Both, the integral as well as the optimization can be
approximated using the available Monte Carlo MCMC samples, by carrying
out the minimization over $\Ab' \in \{\Ab^{(t)};\; t=1,\ldots,T\}$ and by
evaluating the integral as Monte Carlo average.
The posterior point
estimators of other parameters in $\thb$ are obtained as posterior
means conditional on $\widehat{\Ab}$. We evaluate the posterior means
using the posterior Monte Carlo samples. 

\bch
The described MCMC simulation is practicable up to moderately large
$n$, including $n=1000$ in the motivating EHR application. For larger
sample sizes, different posterior simulation methods are needed. We
briefly discuss some suggestions in Section 7, and in Supplementary
Material A. \ech


\section{Simulation Study}
\label{sec:ss}
\begin{figure}
	\centering
	\subfigure[Posterior distribution of number of diseases]{\includegraphics[width=.7\textwidth,height=.15\textheight]{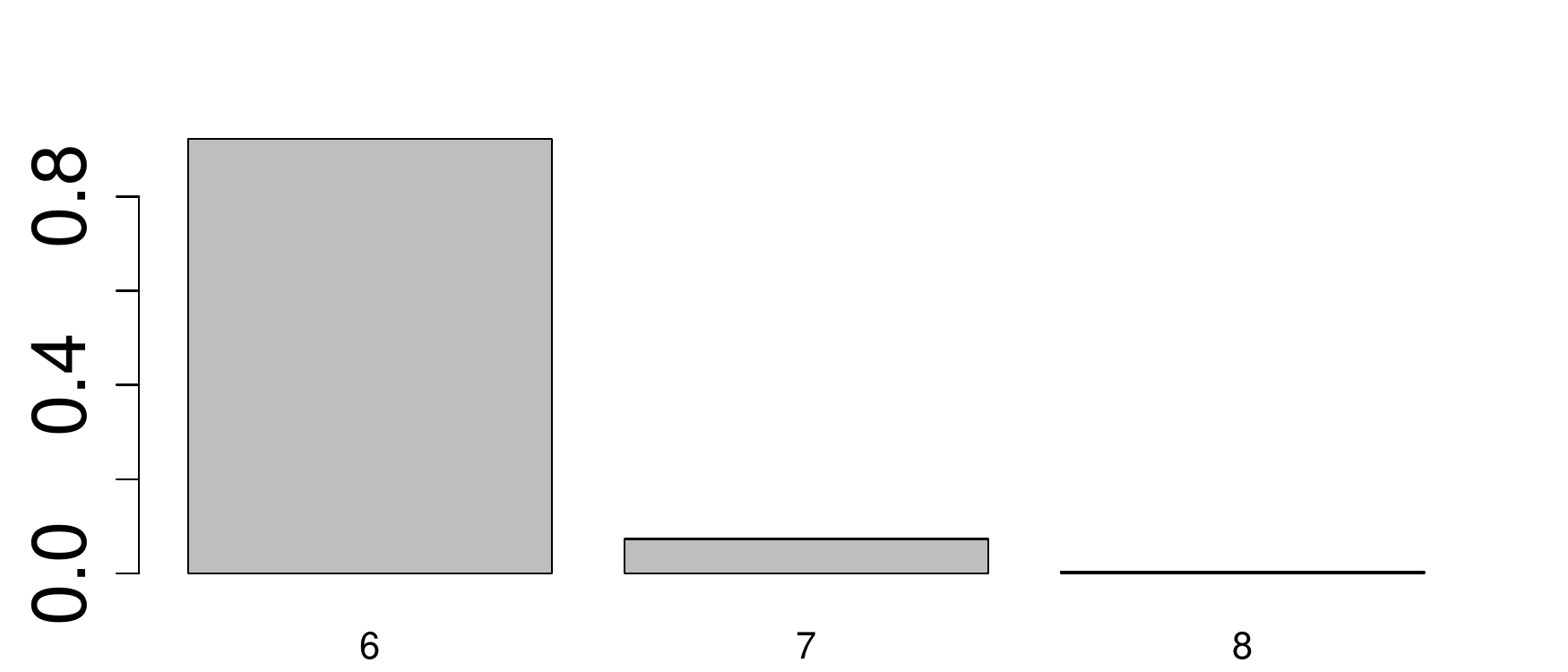}\label{fnde}}
	\subfigure[True $\Ab$]{\includegraphics[width=0.32\textwidth]{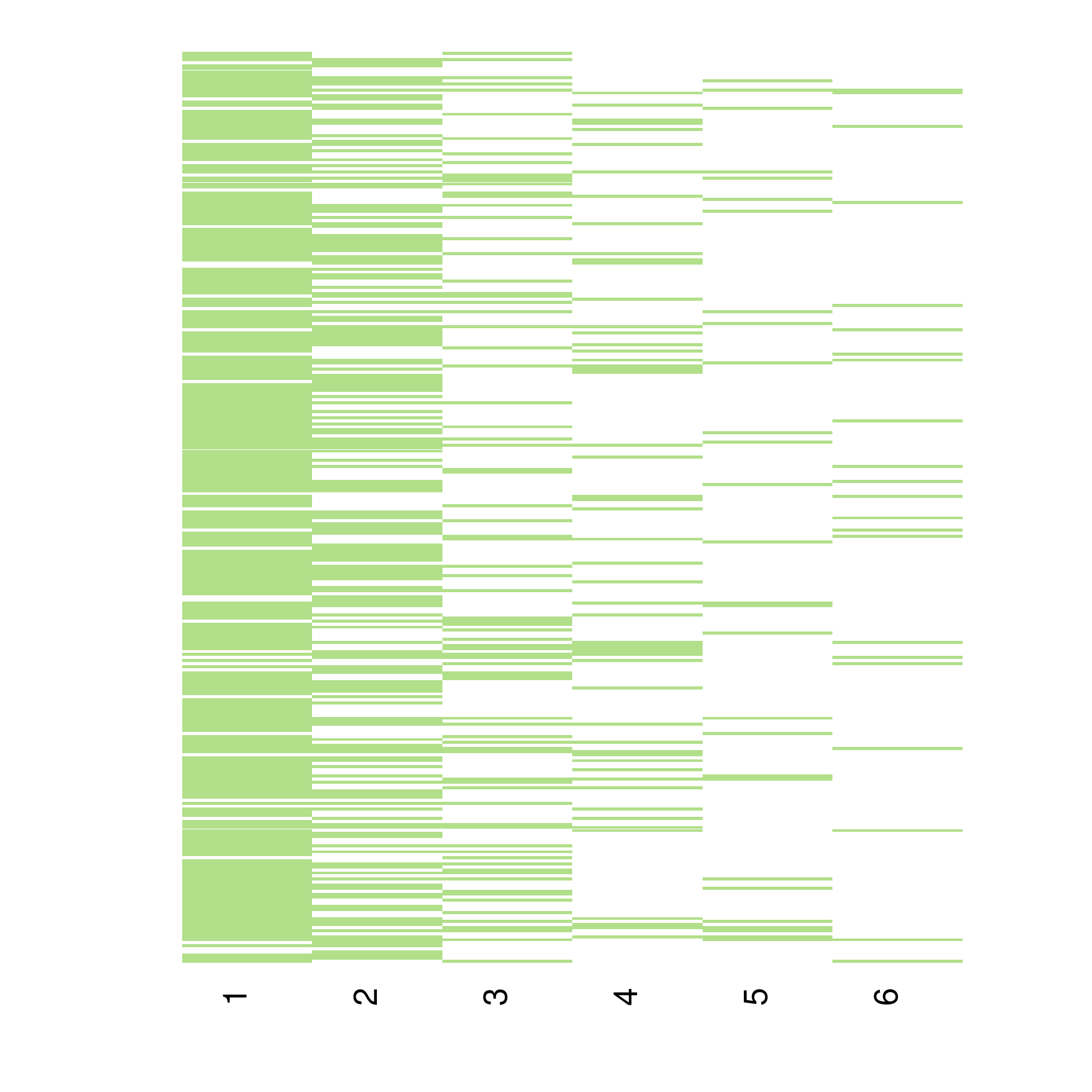}\label{fat}}
	\subfigure[True $\Bb$]{\includegraphics[width=0.32\textwidth]{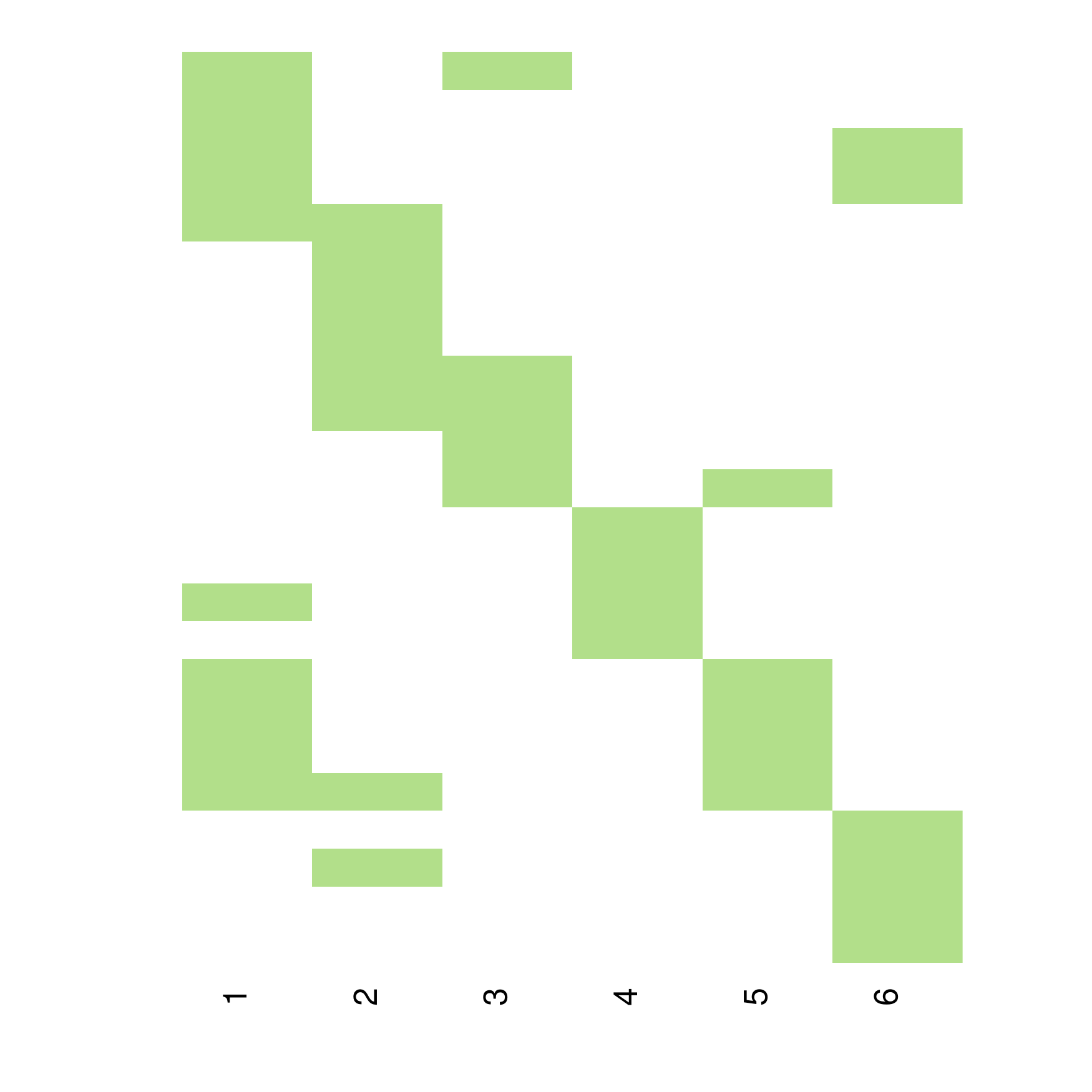}\label{fbt}}	
	\subfigure[True $\Cb$]{\includegraphics[width=0.32\textwidth]{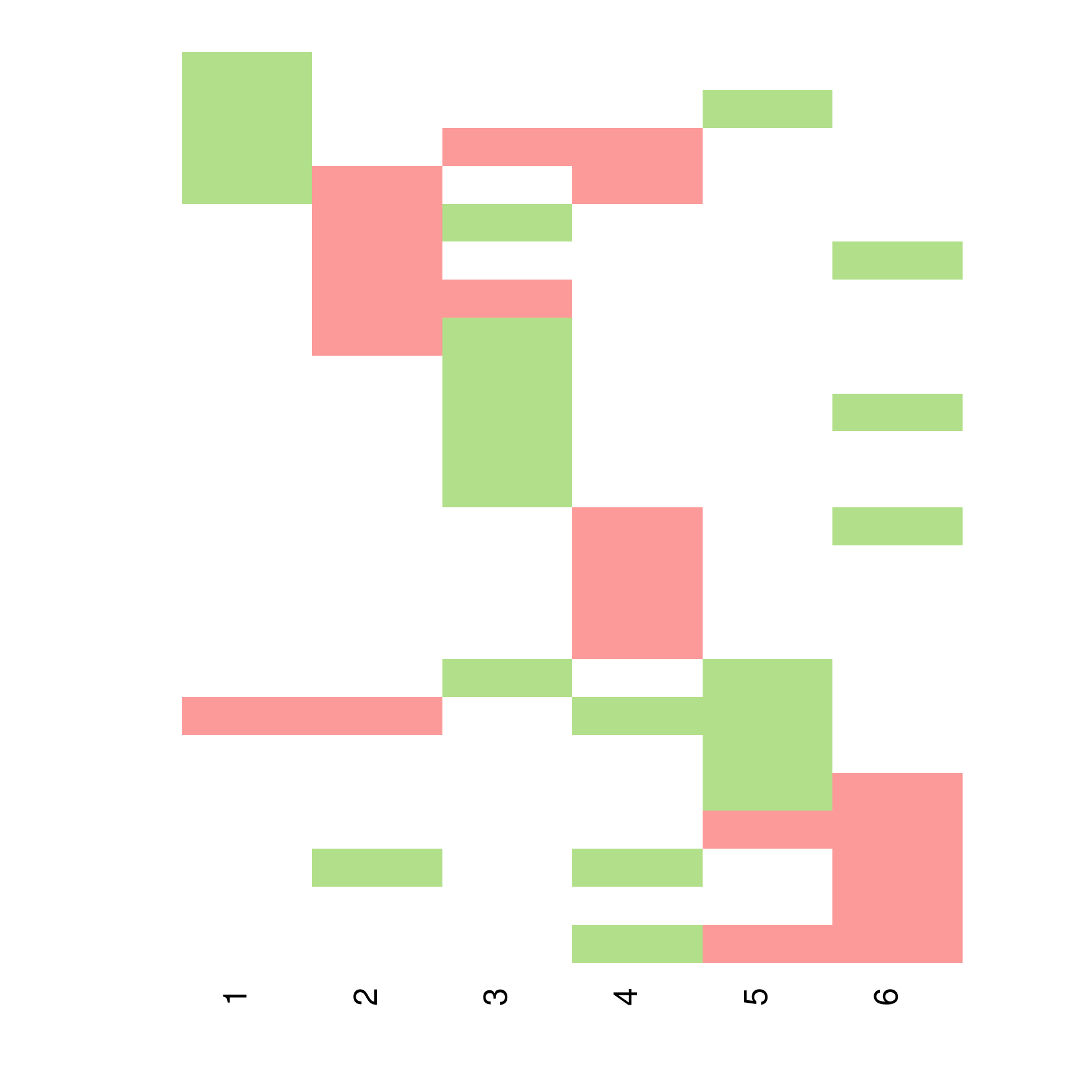}\label{fct}}
	\subfigure[Estimated $\widehat{\Ab}$]{\includegraphics[width=0.32\textwidth]{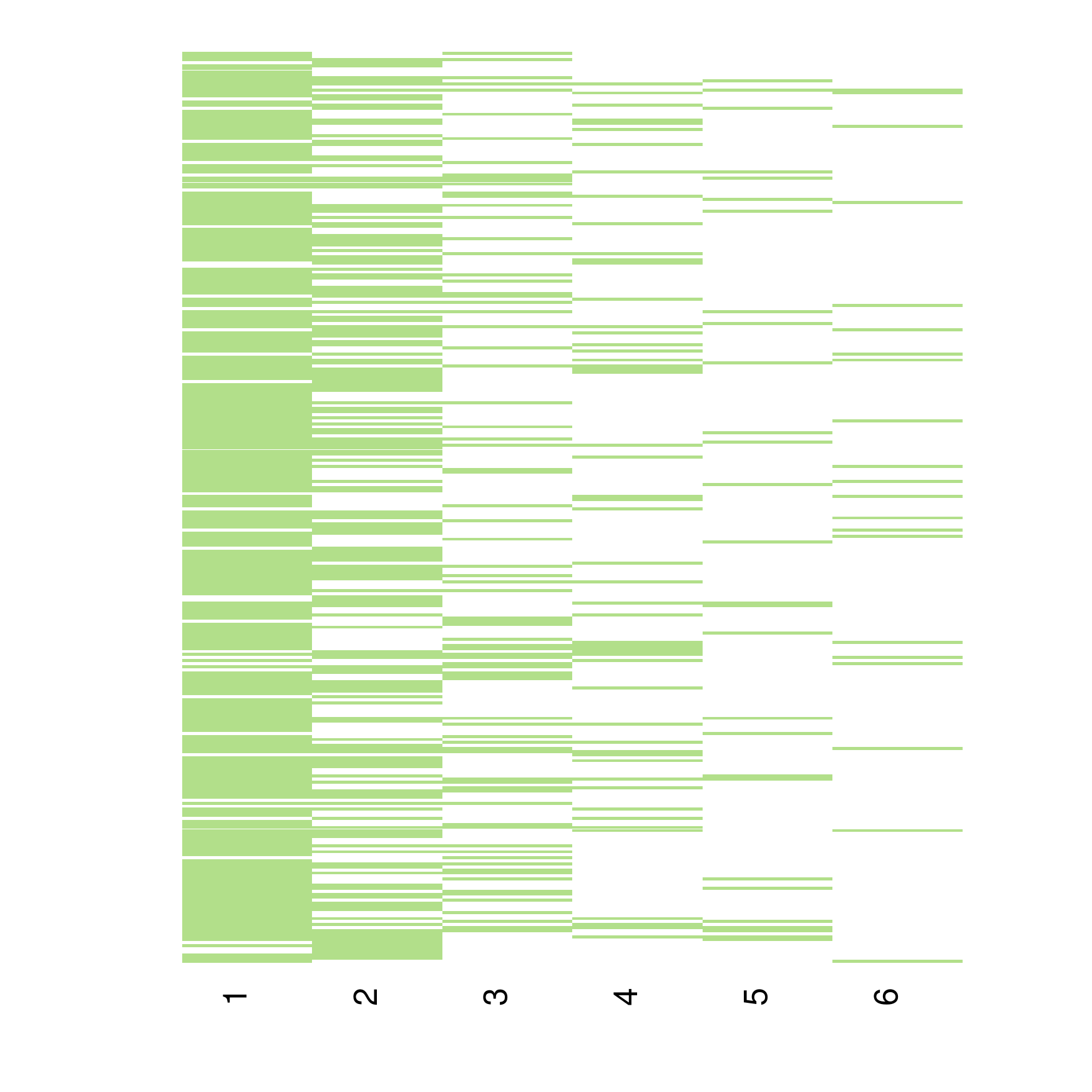}\label{fae}}
	\subfigure[Estimated $\widehat{\Bb}$]{\includegraphics[width=0.32\textwidth]{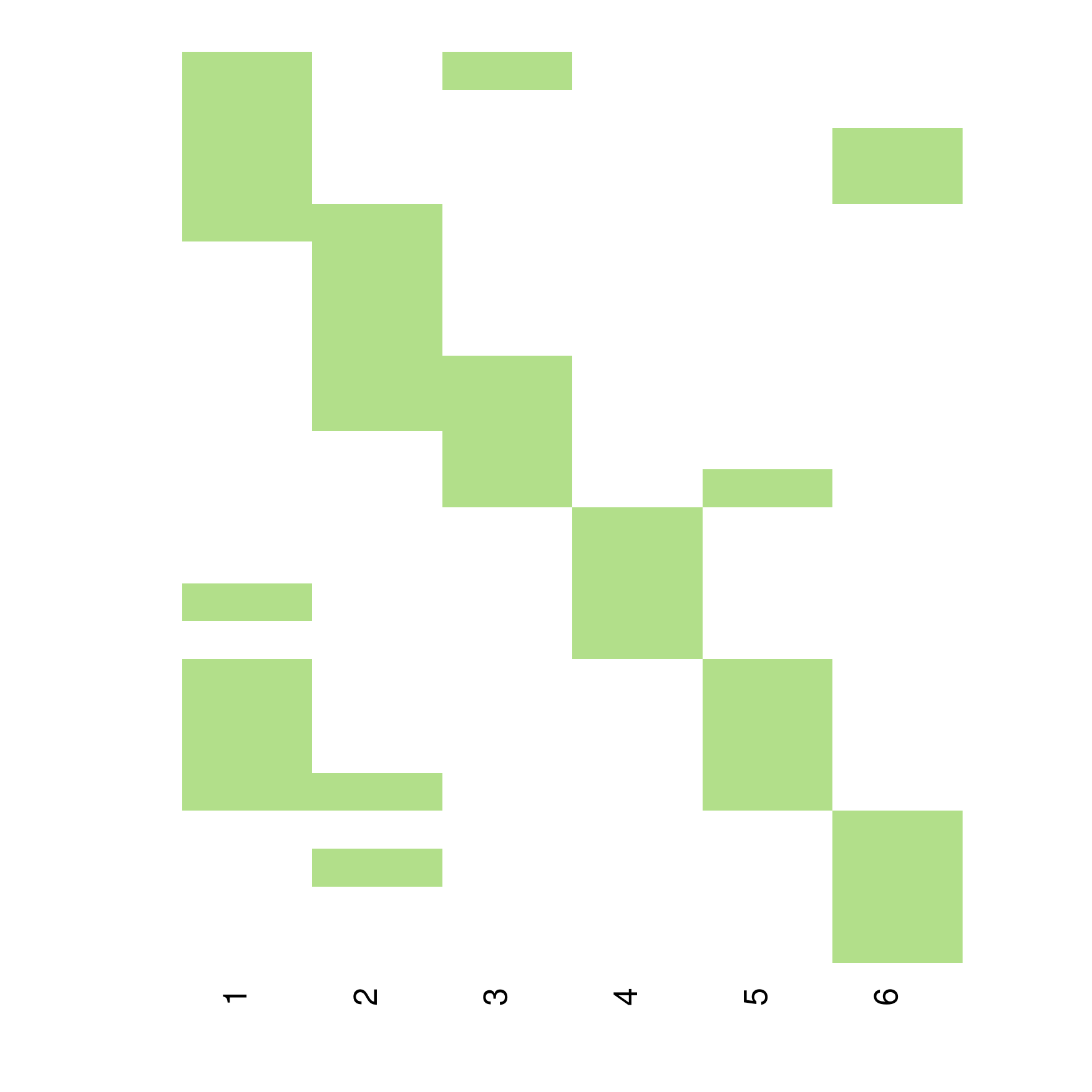}\label{fbe}}	
	\subfigure[Estimated $\widehat{\Cb}$]{\includegraphics[width=0.32\textwidth]{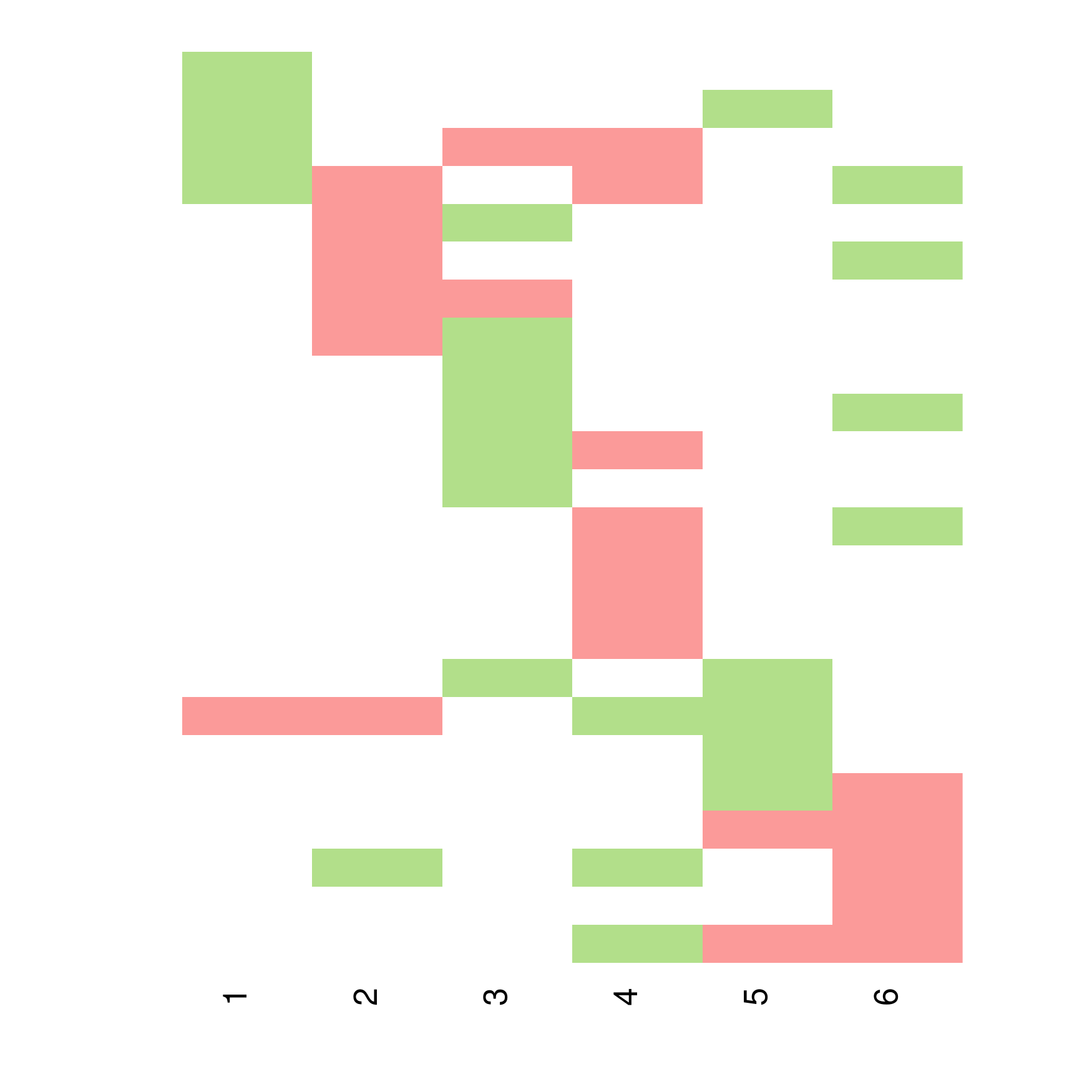}\label{fce}}
	\subfigure[$\widehat{K}$ across 50 simulations.]{\includegraphics[width=0.7\textwidth,height=.15\textheight]{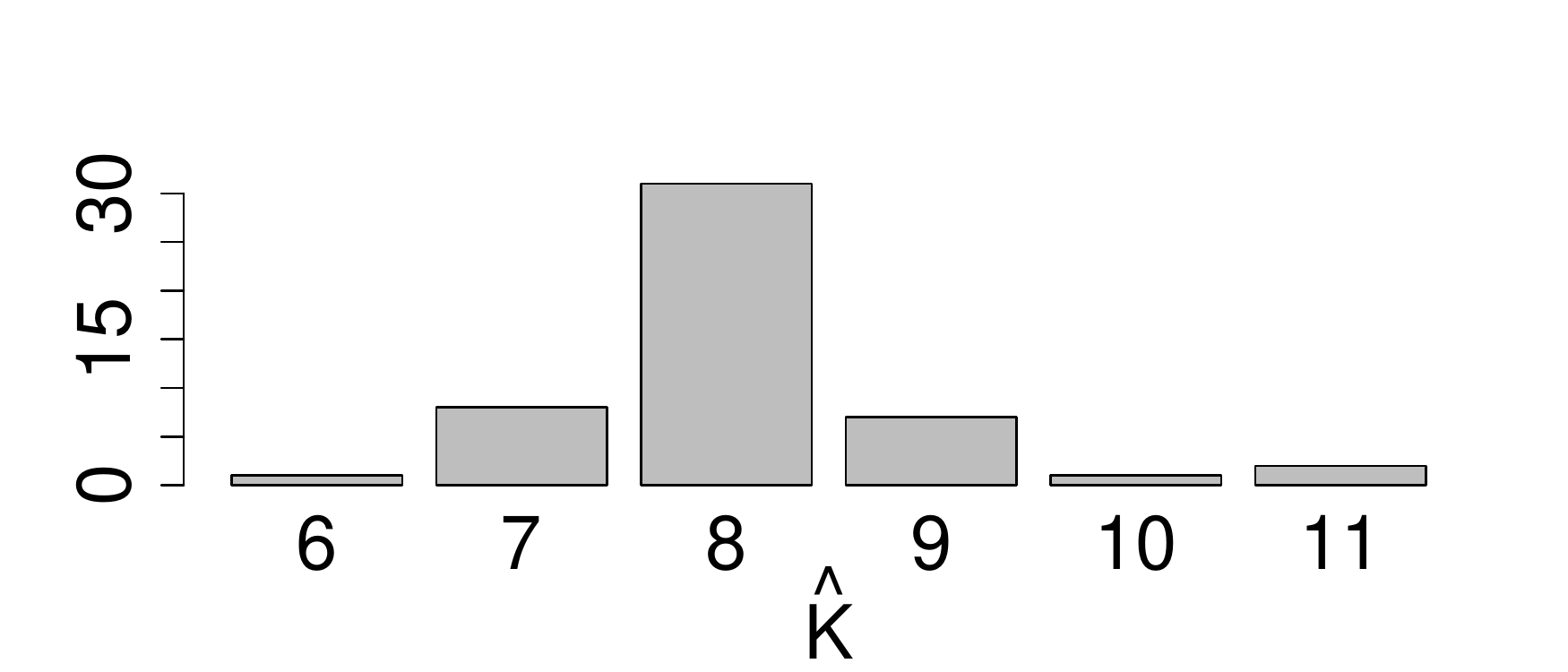}\label{fig:khat_barplot}}
	\caption{Simulation truth and posterior estimates. Panels (a)-(g) are from Scenario I and Panel (h) is from Scenario II. In the heatmaps, green cells represent 1, white cells 0 and red cells -1.  }\label{fig:simre}	
\end{figure}

We consider two simulation scenarios. In both scenarios, we generate
the patient-disease matrix $\Ab$ from an IBP($m$) model with $m=1$ and
sample size $n=300$.
%
The resulting matrix $\Ab$ has $K=6$ columns and
$n=300$ rows, displayed in Figure \ref{fat}. Given $K=6$, we generate
a binary symptom-disease matrix $\Bb\in \{0,1\}^{p\times K}$ and a
categorical symptom-disease matrix $\Cb\in \{-1,0,1\}^{q\times K}$
with $p=q=24$ in the following manner. We first set $\beta_{jk}=1$ for
$k=1,\dots,6$ and $j=4(k-1)+1,\dots,4k$, $\gamma_{lk}=-(-1)^k$ for
$k=1,\dots,6$ and $l=4(k-1)+1,\dots,4k$. We then randomly change 10\%
of the zero entries in $\Bb$ to 1 and 10\% of the zero entries in
$\Cb$ to  +1 or -1. The resulting matrices $\Bb$ and $\Cb$ are shown
in Figures \ref{fbt} and $\ref{fct}$. 
\smallskip

In \underline{Scenario I}, the observations $\Zb$ and $\Yb$ are
generated from the sampling model i.e. equations (\ref{eqn:zij}) and
($\ref{eqn:cat}$). To mimic the  Chinese EHR data, we assume that we
have diagnoses for the first disease and that we know the related
symptoms for the first disease. In the model, we therefore fix the
first column of $\Ab$, $\Bb$ and $\Cb$ to the truth. In addition, we
assume that we have partial information about the second latent disease:
the symptom-disease relationships are known, but no diagnostic
information is available. Accordingly, we will fix the second columns
of $\Bb$ and $\Cb$, but leave the second column of $\Ab$ as unknown
parameters. We ran the MCMC algorithm described in Section
\ref{sec:pi} for 5,000 iterations, \bch which took  $<5$ minutes on a
desktop computer with a 3.5 GHz Intel Core i7 processor. \ech The first
half of the iterations are discarded as burn-in and posterior samples
are retained at every 5th iteration afterwards.  

Inference summaries are reported in Figure \ref{fig:simre}. Figure
\ref{fnde} shows the posterior distribution of the number $K$ of
latent diseases. The posterior mode occurs at the true value
$\widehat{K}=6$. Conditional on $\widehat{K}$, the posterior point
estimate $\widehat{\Ab}$ is 
displayed in Figure \ref{fae} with mis-allocation rate
$\mathcal{H}(\widehat{\Ab},\Ab)/(n\cdot K)=3\%$\footnote{The
  percentage is computed based on free parameters in $\Ab$
  only.}. Conditional on $\widehat{\Ab}$, the point 
estimates $\widehat{\Bb}$ and $\widehat{\Cb}$ are provided in Figures
\ref{fbe} and \ref{fce}. 
The similarity between the heatmaps of
the simulation truth and estimates indicates an overall good recovery of
the signal.
The error rates in estimating $\Bb$
and $\Cb$ are 0\% and 2\%, respectively. We repeat this
simulation 50 times. In 96\% of the repeat simulations, we
correctly identify the number 
$K$ of latent diseases; in the remaining 4\%, it is overestimated by
1. When $K$ is correctly estimated, the average mis-allocation rate,
error rates for $\Bb$ and $\Cb$ are 3\%, 1\% and 1\% with standard
deviation 0.5\%, 1\% and 1\%, respectively. \bch We provide addtional
simulation studies in Supplementary Material A to investigate the
performance of DFA with different hyperparameters and with the
alternative prior that was introduced in \eqref{eqn:hprior}.  \ech
\smallskip

In \underline{Scenario II}, we use a different simulation truth 
and generate data $\{\Zbt,\Ybt\}$ from latent factor models 
\begin{align*}
  \Zbt=\Phib\Lambdab_z+\Eb_z\mbox{~~and~~}\Ybt=\Phib\Lambdab_y+\Eb_y
\end{align*} 
with latent factor matrix $\Phib=\Ab\circ \widetilde{\Phib}$ and
loading matrices $\Lambdab_z=\Bb\circ  \widetilde{\Lambdab}_z,
\Lambdab_y=\Cb\circ  \widetilde{\Lambdab}_y$ where $\Ab,\Bb,\Cb$ are
the same as in Scenario I.  The elements of $\widetilde{\Phib},
\widetilde{\Lambdab}_z, \widetilde{\Lambdab}_y$ are i.i.d. $Unif(0,4)$
and the errors are i.i.d. standard normal. We then threshold $\Zbt$
and $\Ybt$ to obtain $\Zb$ and $\Yb$ at different levels $t>0$: 
\[ 
z_{ij}=\left\{
  \begin{array}{lcl}1 &\mbox{if}&
    \tilde{z}_{ij}>t\\0&\mbox{if}&\tilde{z}_{ij}\leq t
  \end{array}\right.\mbox{~~and~~}
y_{il}=\left\{
  \begin{array}{lcl}
    +1 &\mbox{if}& \tilde{y}_{il}>t\\
    0  &\mbox{if}&|\tilde{y}_{il}|\leq t\\
    -1 &\mbox{if}&\tilde{y}_{il}< -t.
  \end{array}\right.
\] 
We applied DFA to $\{\Zb,\Yb\}$  using different values of the
threshold $t\in\{1,2,3,4,5\}$.
 Also, we include no known diseases. 
Performance deteriorates as $t$ grows. DFA tends to overestimate the
number $\widehat{K}$ of latent factors by 1 to 3 extra factors. After
removing those extra factors, the mis-allocation rate
$\mathcal{H}(\widehat{\Ab},\Ab)/(n\cdot K)$ is between 9\% and
19\%. And the error rates for $\widehat{\Bb}$ and $\widehat{\Cb}$ are
between 1\% and 17\%.

For comparison, we implemented inference also under
the sparse latent factor model (SLFM,
\citealt{rovckova2016fast}) to $\{\Zbt,\Ybt\}$. SLFM assumes a sparse
loading matrix and unstructured latent factors. Therefore, we only
report performance in recovering the sparse structure $\Bb$ and
$\Cb$ of the loading matrices $\Lambdab_z$ and $\Lambdab_y$. The
penalty parameter $\lambda_0$ of SLFM is chosen in an ``oracle" way:
we fit SLFM with a range of $\lambda_0$ values and select $\lambda_0$
that yields the best performance given the simulation truth. The
resulting error rates in estimating $\Bb$ and $\Cb$ are 11\% and 10\%,
comparable to those of DFA  (keeping in mind the oracle choice of
$\lambda_0$). 

We repeat the experiment 50 times for $t=3$. The estimated
$\widehat{K}$ across 50 simulations are plotted in Figure
\ref{fig:khat_barplot}. We observe that DFA tends to overestimate $K$
when model is misspecified. We report the performance of estimating
$\widehat{\Bb}$ and $\widehat{\Cb}$ based on the best
subset of columns.
The mean (standard deviation) error rates in estimating $\Bb$
and $\Cb$ are 8\% (4\%) and 9\% (3\%), respectively. SLFM has slightly
higher error rates, 12\% (1\%) and 11\% (1\%).


\section{Phenotype Discovery with EHR Data}
\label{sec:cs}

\subsection{Data and preprocessing}

We  implement latent disease mining for 
the EHR data introduced in Section \ref{sec:mcs}.
Using the reference range 
for each test item, we define a symptom if the value of an item
falls beyond the reference range. Some symptoms are binary in nature,
e.g. low density lipoprotein is clinically relevant only when it is
higher than normal range. Other symptoms are inherently ternary,
e.g. heart rate is symptomatic when it is too high or too low. 
The 39  testing items are listed  in Table \ref{tab:names} where we
also indicate  which items give rise to a binary or ternary
symptom.  

We extract diagnostic codes for diabetes from the sections ``medical
history" and ``other current diseases" in the physical examination
form. A subject is 
considered as having diabetes if it is listed in either of the two
sections. There are 36 patients diagnosed with diabetes. We fix the
first column of $\Ab$ in the PD model according to the diabetes
diagnosis. Moreover, it is known that diabetes is clinically
associated with high glucose level. We incorporate this prior
information by fixing the corresponding entry in the first column of
$\Cb$ in the SD model.

There is additional prior knowledge about symptom-disease
relationships. 
Creatinine, a waste product from
muscle metabolism, is controlled by the kidneys to stay within a
normal range. 
Creatinine has therefore been found to be a reliable indicator of kidney
function. Elevated creatinine level suggests damaged kidney function
or kidney disease. Blood urea nitrogen (BUN) level is another
indicator of kidney function. Like creatinine, urea is also a
metabolic byproduct which can build up if kidney function is
impaired. We fix the two entries (corresponding to creatinine and BUN)
of the second column of $\Cb$ to 1 and the rest to 0. With this prior
knowledge, we interpret the second latent disease as \textit{kidney
  disease}. Likewise, it is known that elevated systolic blood
pressure and diastolic blood pressure are indicators of
hypertension, and abnormal levels of total bilirubin (TB), aspartate aminotransferase (AST) and alanine aminotransferase (ALT) are indicators of liver diseases. We fix the corresponding entries of the third and fourth column of
$\Bb$ and $\Cb$,
and interpret the third latent disease as \textit{hypertension} and the fourth latent disease as \textit{liver disease}.

To comply with Chinese policy, we report inference for data preprocessed
by a Generative Adversarial Network (GAN,
\citealt{goodfellow2014generative}), which replicates  the
distribution underlying the raw data. 
GAN is a machine learning algorithm which
simultaneously trains a generative model and a discriminative model on
a training dataset (in our case, the raw EHR dataset). The generative
model simulates a hypothetical repetition of the 
training data, which is then combined with the original training data to
form a merged data set.
Meanwhile, the discriminative model tries to distinguish between
original data and simulations in the merged data set.
During training, the generative model uses gradient information from
the discriminative model to produce better simulations.
 Training continues until the discriminative model can no longer
distinguish. 
After training, the generative
model can be used to generate an arbitrary number of simulations
which are similar in distribution to the original dataset. Any statistical inference in the original data and the replicated
data is identical to the extent to which it relies on low
dimensional marginal distributions. 
\bch In our case, we generate a simulated dataset of the same size as the
raw EHR dataset and then discretize it using the reference range. \ech A similar approach has been used in
\cite{ni2018scalable}. 

\subsection{Results and interpretations}
We ran the MCMC algorithm described in Section \ref{sec:pi} for 50,000
iterations. The first half of the iterations are discarded as burn-in
and posterior samples are retained at every 5th iteration thereafter.
\bch Goodness-of-fit and MCMC convergence diagnostics
 show adequate fit and no evidence for lack of
 convergence (Supplementary Material B). \ech
Posterior probabilities for the number $K$ of latent diseases are
 $p(K=14 \mid data)=0.69$
and $p(K=15 \mid data)=0.24$, respectively, i.e., the maximum a posteriori (MAP)
estimate is $\Khat=14$.
This  includes the 4 
\textit{a priori} known diseases as well as 10 newly discovered
latent diseases. 

\begin{figure}
  \centering
  \includegraphics[width=.4\textwidth]{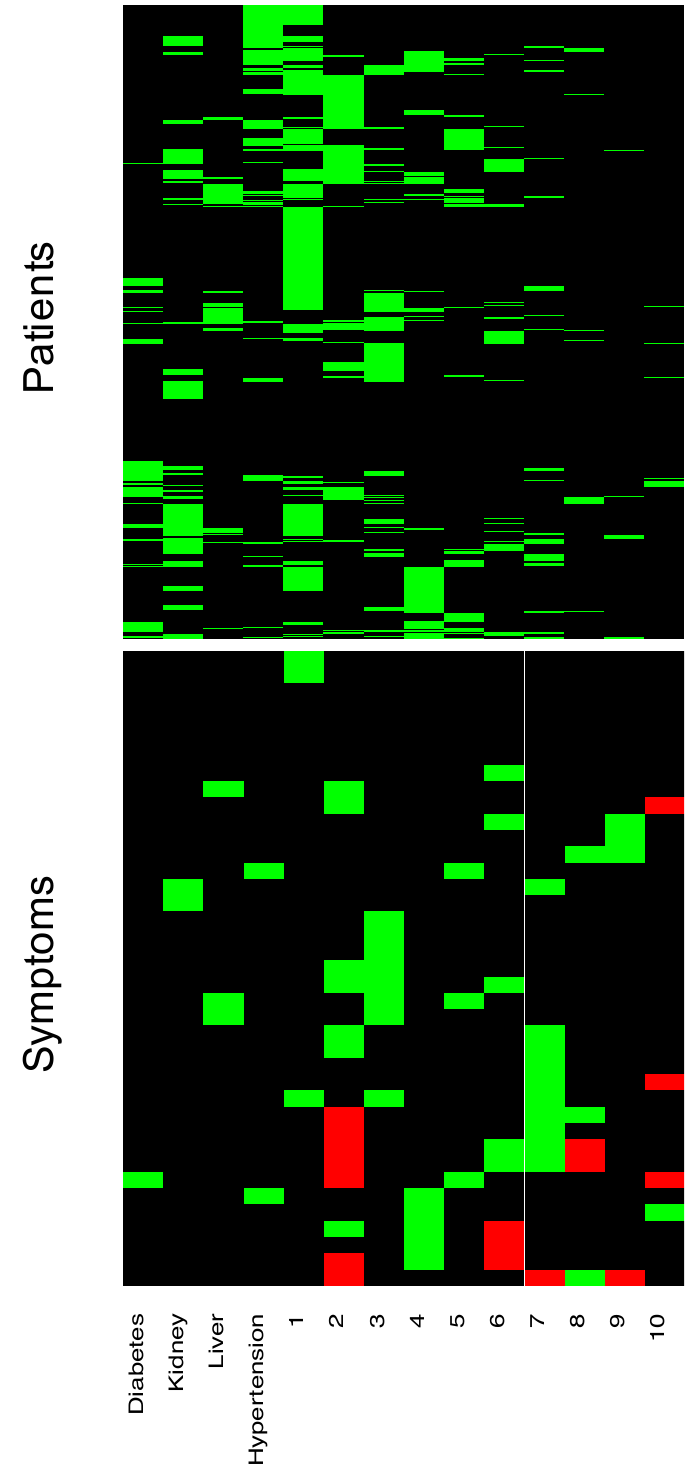}
  \caption{EHR data. The heatmap on top shows the estimated
    patient-disease relationships $\widehat{\Ab}$. The bottom part of
    the double heatmap shows the estimated symptom-disease
    relationships $\widehat{\Bb},\widehat{\Cb}$ with green,
    black and red cells representing 1, 0 and -1,
    respectively. The columns correspond to diseases and the rows are
    patients (top portion) and symptoms (bottom portion).} 
  \label{fig:hmpd}
\end{figure}

Conditional on $\Khat$, the posterior estimates of the PD and SD models
are shown as a heatmap in Figure \ref{fig:hmpd} with green, black
and red cells representing 1, 0 and -1, respectively.
 The nature of the figure as a single heatmap with two blocks, for
PD and SD, respectively, highlights again the nature of the model as a
``double'' feature allocation with matching subsets of patients
and symptoms. 
The model allocates both, patients and symptoms, to latent diseases.
As mentioned before,
DFA can also be interpreted as an edge-labeled
network. We show the same results as in the heatmap as 
a bipartite graph in Figure \ref{fig:net_result}. The full inferred
model  would be a tripartite network, as
in the bottom portion of Figure \ref{fig:net}. 
However, we omit the patient nodes in Figure \ref{fig:net_result} ,
lest the figure would be overwhelmed by the patient nodes.
Instead,
we  summarize the patient-disease relationships by specifying the font
size of the disease node (triangle, purple font) proportional to 
the number of linked patients.  Latent diseases are
labeled  by numbers and a priori known diseases are labeled by
name. Symptoms are shown in black font, with
lines showing the links to diseases. Dashed lines
are symptom-disease relationships that are inferred from the data whereas solid
lines are fixed by prior knowledge. Black lines indicate that symptoms
are binary. Red (blue) lines indicate suppression (enhancement) relative to the 
normal range. Line widths are proportional
to the posterior probabilities of edge inclusion.

We find  239  patients with impaired kidney function or kidney
disease, 183 patients with hypertension and 93 patients with liver
disease. The prevalence of kidney disease is slightly higher than the
national average 16.9\% \citep{zhang2012prevalence} probably because
of the elderly patient population in this study. \bch We caution that the estimated prevalence from our analysis should be viewed as an estimate for the lower bound of the actual prevalence in the target patient population because we are not explicitly addressing confounders such as treatment. For example, our estimated hypertension
prevalence (18.3\%) is much lower than the national average 57.3\%
\citep{zhang2017hypertension},  probably simply due to successful
and widely available treatment. 
Individuals who were previously diagnosed with hypertension and comply
with the prescribed medication may not show any symptom (high blood
pressure) in the physical examination. \ech

We identify additional 10 latent diseases with prevalence of 493, 218,
192, 174, 114, 82, 64, 24, 15,  and 15 patients. Some of the latent
diseases are quite interesting. 
\underline{Latent disease 1} is \textit{lipid disorder},
associated with high total cholesterol (TC), triglycerides and low
density lipoprotein (LDL). Cholesterol is an organic molecule carried
by lipoproteins. LDL is one type of such lipoproteins, commonly
referred to as ``bad" cholesterol. At normal levels, TC and LDL are
essential substances for the body. However, high levels of TC and LDL
put patients  at increased risk for developing heart disease and
stroke. Triglycerides are a type of fat found in the blood which are
produced by the body from excessive carbohydrates and fats. Like
cholesterol, triglycerides are essential to life at normal
levels. However, a high level is associated with an increased chance for
heart disease.

\underline{Latent disease 3}
\bch can be characterized as  \ech
\textit{thrombocytopenia-like} disease which causes low
count of platelets, decreased plateletcrit (PCT) and coefficient of variation of platelet distribution (PDW-CV), and increased mean platelet volume (MPV). Patients with low
platelets may not be able to stop bleeding after injury. In more
serious cases, patients may bleed internally which is a
life-threatening condition. 

\underline{Latent disease 4} is a
\textit{polycythemia-like} disease, associated with elevated mean corpuscular hemoglobin concentration (MCHC), hemoglobin, erythrocytes and hematocrit
(HCT). These symptoms match exactly  the symptoms of 
polycythemia,
a disease that gives rise to an increased level of circulating red
blood cells in the bloodstream. Polycythemia can be caused
intrinsically by abnormalities in red blood cell production or by
external factors such as chronic heart diseases. 

Interestingly, like latent disease 4, \underline{latent disease 6} is
also related to hemoglobin, erythrocytes and HCT. However, it is
linked with a decrease in these levels; hence we refer
to the disease as \textit{anemia}. 

\underline{Latent disease 7} suggests
\textit{bacterial infection} with increased  leukocytes, granulocytes (GRA) and heart rate, and decreased monocytes (MON) and lymphocytes (LYM). The immune system, specifically the bone marrow, produces more GRA and leukocytes to fight a bacterial infection. As a result, the relative abundance of MON and LYM decreases.

Relatedly, \underline{latent disease 8} may be caused by \textit{viral
  infection}. Viruses can disrupt the  function  of bone marrow which leads to
low levels of leukocytes and GRA, and high levels of LYM.  

\underline{Latent disease 9} is related to \textit{allergy} with abnormal basophil, GRA and LYM. 

\underline{Latent disease 10} suggests that a small group of patients
may have \textit{malnutrition}, which  is linked with  low blood glucose and
anemia-like symptoms such as low corpuscular volume (MCV) and
corpuscular hemoglobin (MCH). 

Such automatically generated inference on latent disease phenotypes is
of high value for routine health exams. It helps the practitioners to
focus resources on specific patients and suggests meaningful
additional reports.
Inference in the statistical model can of course not replace 
clinical judgment and needs further validation. But it can provide an important
decision tool to prioritize resources, especially for areas with limited medical support such as the areas where our data are collected from. 

We remark that although there are clear interpretations for most
  latent diseases found by DFA, latent diseases 2 and 5 cannot
  easily be
  interpreted as  specific diseases.  Latent disease 2 is associated with
  12 symptoms, which is likely beyond the number of symptoms of any single disease.  Most of the symptoms, such as platelets,
  leukocytes and lymphocytes, are due to a weak immune
  system. Considering the elderly population of this dataset, aging
  could be a reasonable cause. Decreased heart rate and
  low glucose level can also be explained by aging. Latent disease 5
  is linked with elevated systolic blood pressure and glucose. While
  those two symptoms may not be simultaneously linked to the same
  disease, their co-appearance should not be too surprising because
  the co-existence of hypertension and diabetes (to which blood
  pressure and glucose are known to be linked) is quite common
  \citep{de2017diabetes}. 
  
\bch We report the estimation of $\{\Wb_j,\zeta_j\}_{j=1}^p$ and
$\{\Wb^-_l,\Wb^+_l,\eta_l^-,\eta_l^+\}_{l=1}^q$ in Supplementary
Material B. The  estimated  baseline weights $\zeta_j, \eta_l^-,
\eta_l^+$ are significantly smaller than disease-related weights
$\Wb_j,\Wb^-_l,\Wb^+_l$. \ech 
%

\subsection{ A Web application for disease diagnosis}
\label{sec:shiny}
\begin{figure} 
	\begin{center}
		\subfigure{\fbox{\includegraphics[height=.4\textheight]{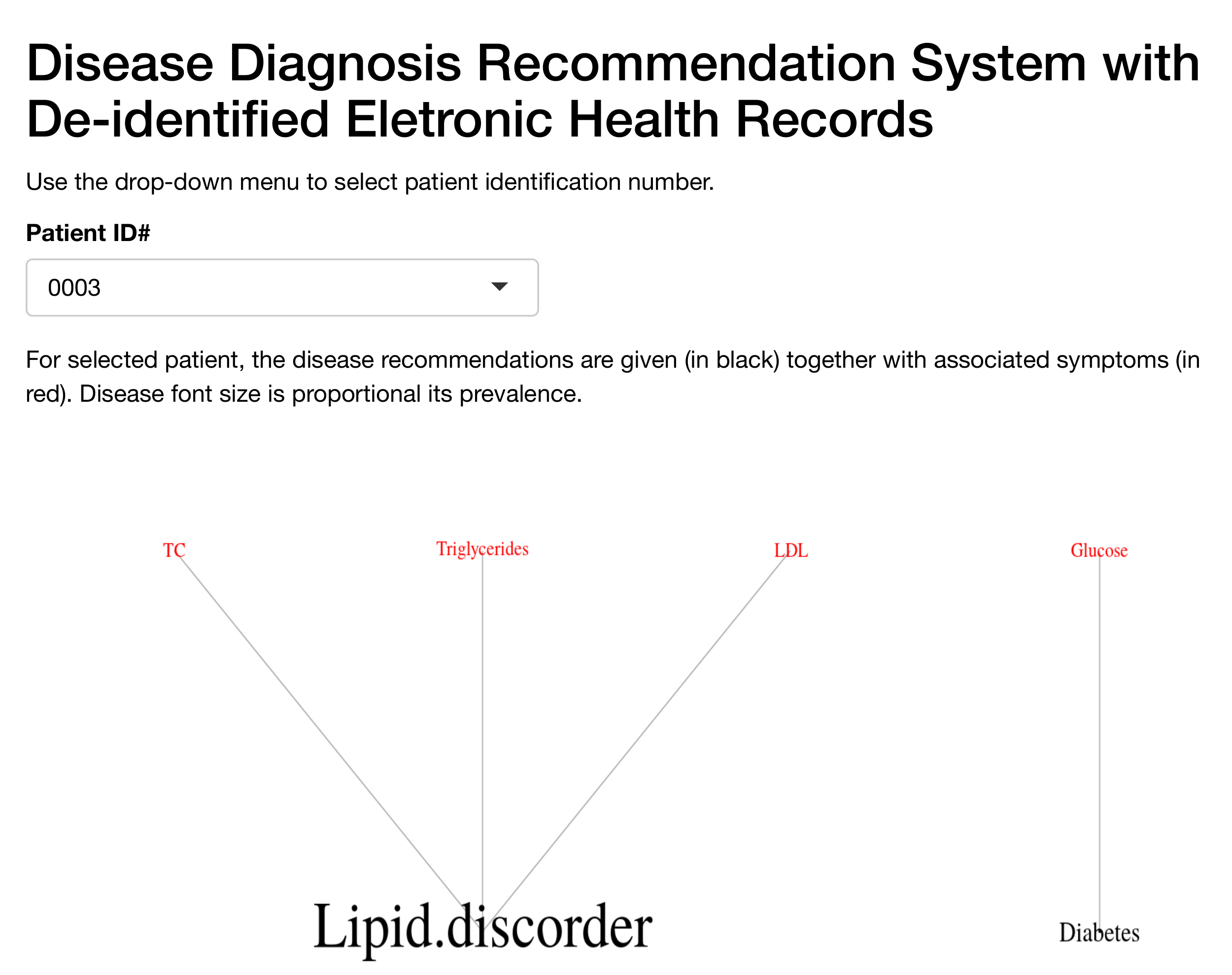}}}
		\subfigure{\fbox{\includegraphics[height=.4\textheight]{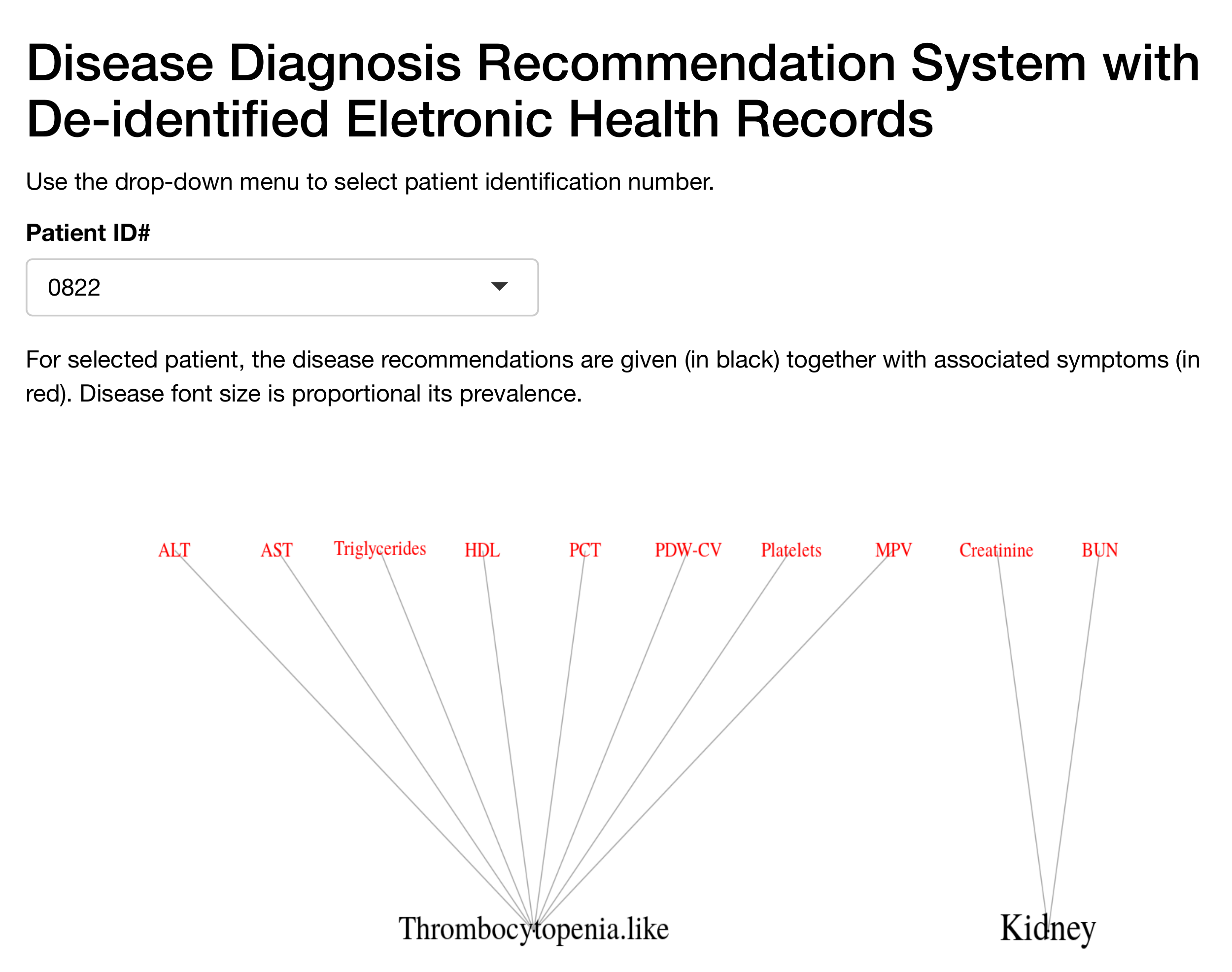}}}
		\caption{Two examples of the R Shiny web application. Each example is for one patient. The application interactively displays disease diagnosis recommendations for de-identified patients selected by the user.}\label{screenshots}
	\end{center}
\end{figure}

A good user interface is critical to facilitate the implementation of
the proposed approach in the decision process of healthcare providers,
and for broad application. Using the R package {\tt shiny}
\citep{RShiny}, we have created an interactive web application
(available at \url{https://nystat3.shinyapps.io/Rshiny/}). The
application displays disease diagnosis recommendations for
de-identified patients selected by the user. We show the application
for two patients in Figure \ref{screenshots}. 

\subsection{Comparison}

\begin{figure}
	\centering
		
	\subfigure[SD bipartite network from DFA]
        {\includegraphics[width=.56\textwidth]{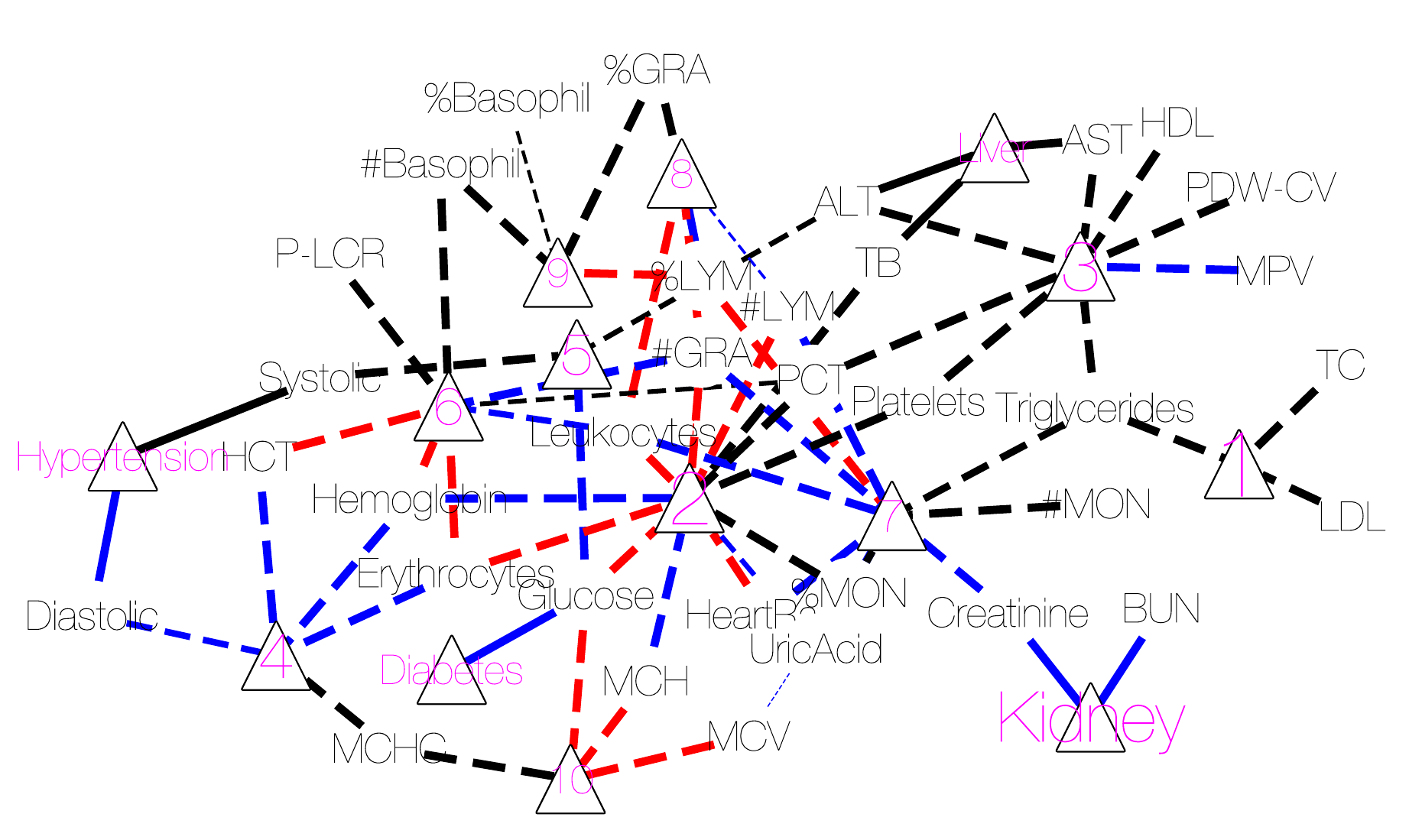}\label{fig:net_result}}
	\subfigure[SD bipartite network from SLFM]
        {\includegraphics[width=0.43\textwidth]{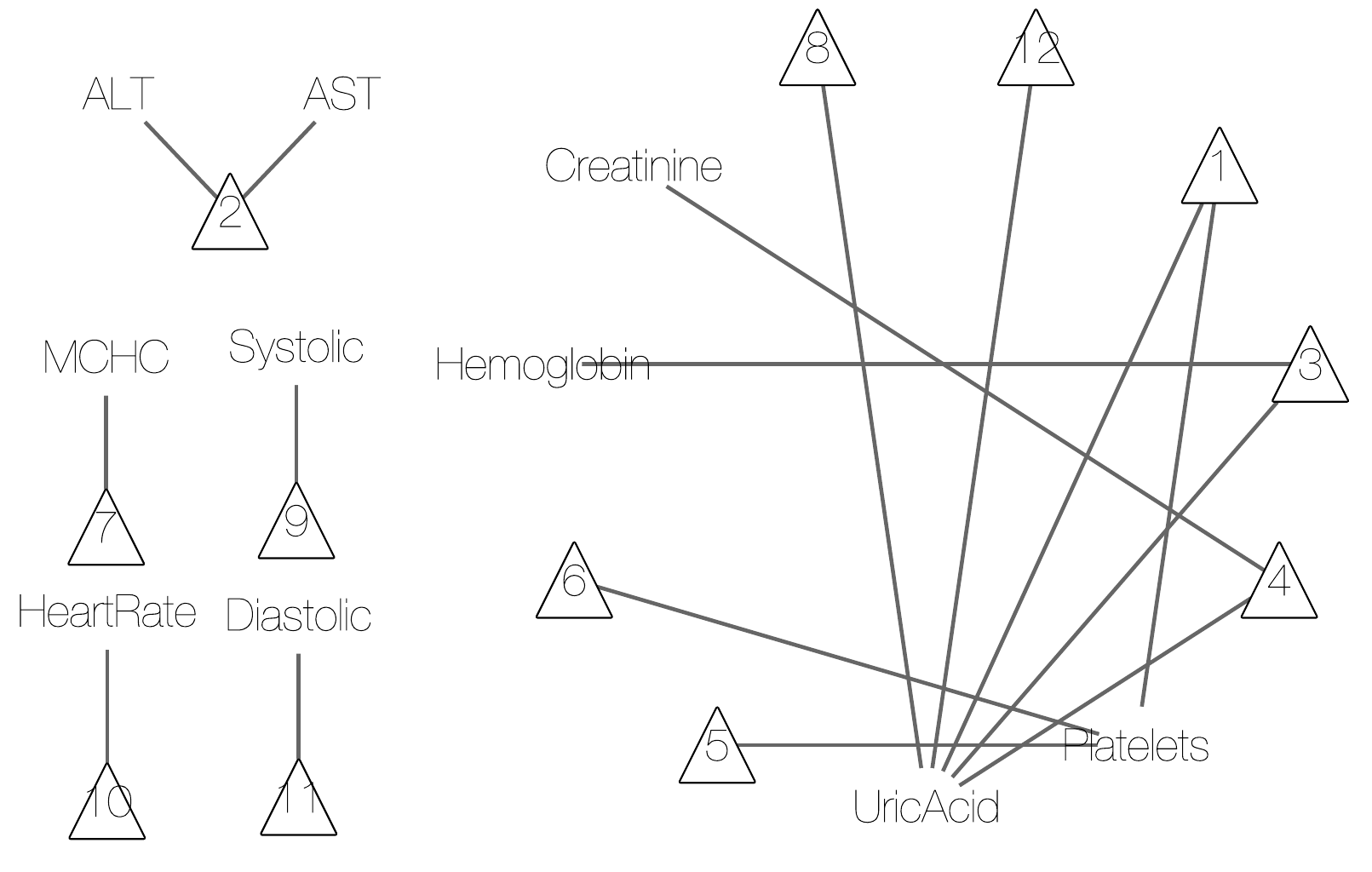}\label{fig:net_LFM2}}	
	\subfigure[SD bipartite network from SLFM with prior
        information]
        {\includegraphics[width=0.56\textwidth]{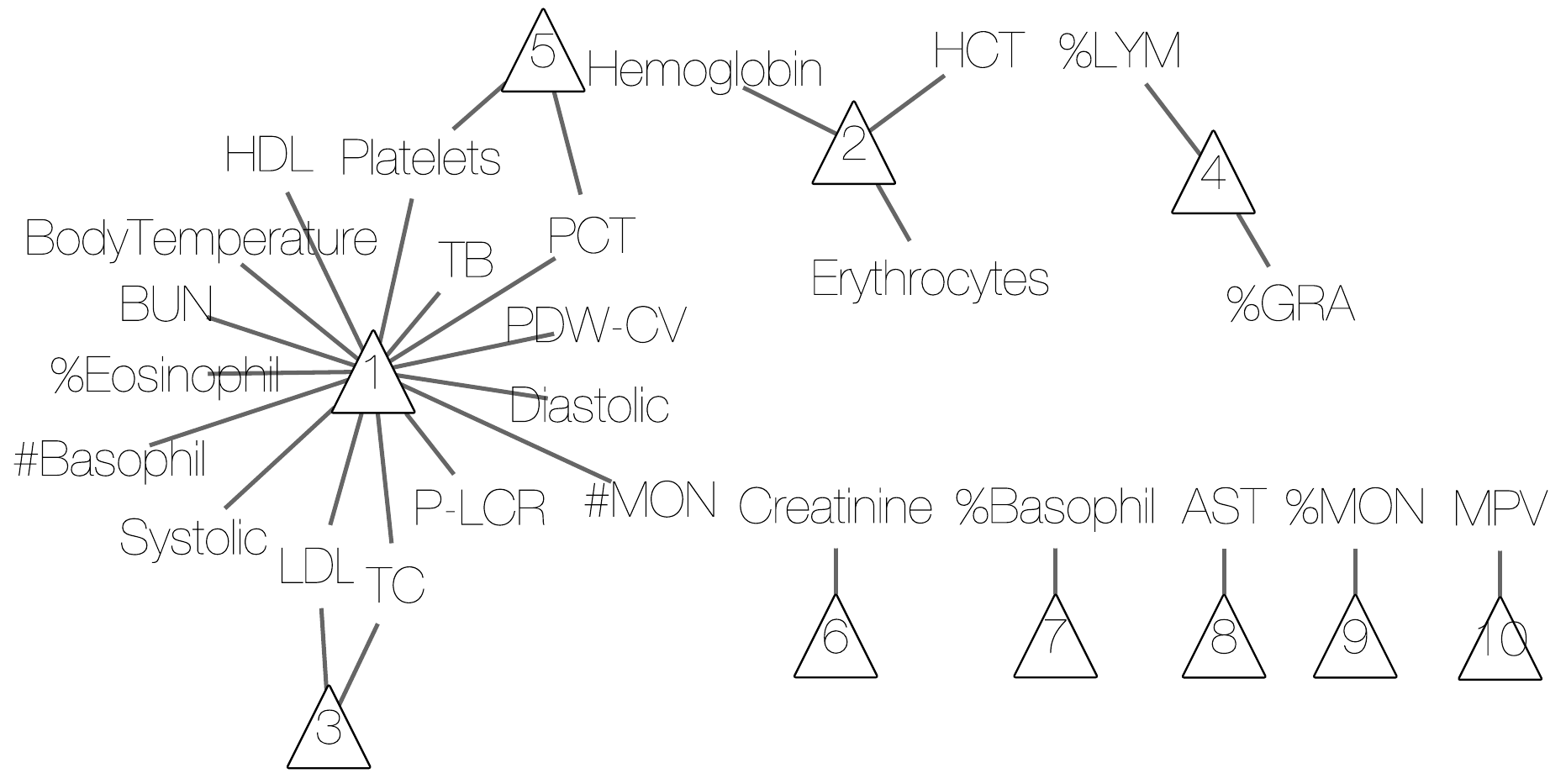}\label{fig:net_LFM_ref}}	
	\subfigure[Symptom network from BDgraph]
        {\includegraphics[width=0.4\textwidth]{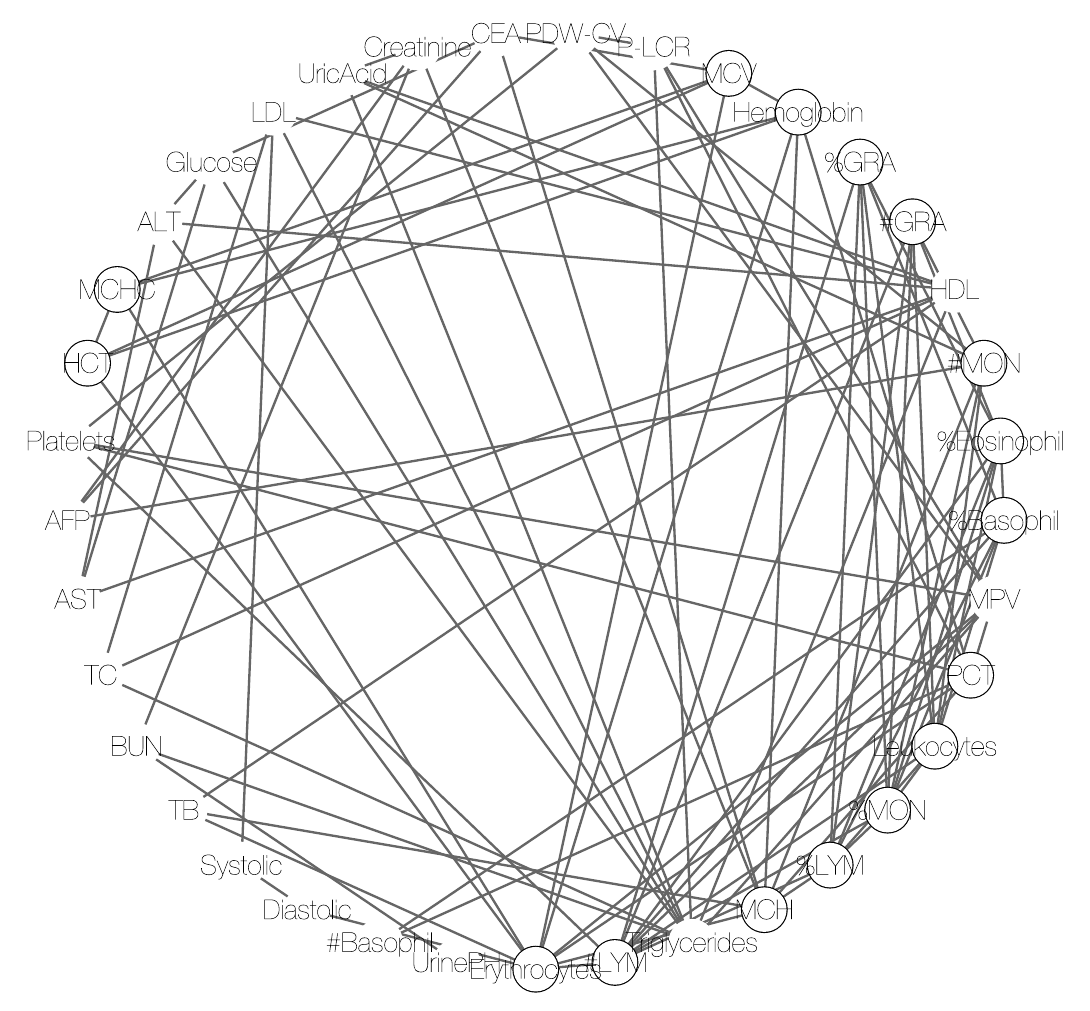}\label{fig:net_BD}}
	\caption{EHR data analysis.  (a) Bipartite network for symptom-disease relationships from DFA. The diseases are represented by triangles with the font size proportional to its popularity (i.e. the number of patients having the disease). Latent diseases are represented by the numbers (10 latent diseases in total). The symptoms are given in black font and their links to each disease are represented by the lines. Dashed lines are symptom-disease relationships inferred from the data whereas solid lines are fixed by prior knowledge. Black lines indicate the symptoms are binary. Red (blue) lines indicate the disease causes the symptom to be lower (higher) than normal range. The line width is proportional to its posterior probability of inclusion. (b) Bipartite network for symptom-disease relationships from SLFM. Latent diseases are represented by triangles (8 latent diseases in total). (c) Network for symptom-disease relationships from BDgraph.  There are 5 cliques of length greater than 3 in total. The symptoms that form those cliques are represented by circles.}\label{fig:read}	
\end{figure}

As a comparison with results under alternative approaches,
we consider inference under SLFM, applied directly to the blood test
results, without converting to binary or ternary symptoms. 
The tuning parameter $\lambda_0$ was set to 1 to
approximately match the number of latent diseases found by DFA. 
SLFM implements inference on sparse symptom-disease relationships as shown in
Figure \ref{fig:net_LFM2}. Although there is no known truth for
symptom-disease relationships, it is difficult to interpret certain
links. While uric acid may play some role in certain diseases, we
do not expect it to be related to 5 out of 12 latent
diseases. In addition, both latent diseases 5 and 6 are related to
platelets only, which should be collapsed into one disease. We also
ran LSFM with larger $\lambda_0$ but found similar results. For
example, when $\lambda_0=10$, SLFM identifies 20 latent diseases, 17
of which are associated with uric acid. These somewhat surprising
results may be due to the assumption of normality and linearity of
SLFM, and taking no advantage of prior information.

\bch Next we apply SLFM to the EHR data that are scaled and centered
at the midpoint of each reference range. Choosing $\lambda_0=13$, the
algorithm found 10 latent diseases whose relationships with symptoms
are depicted in Figure \ref{fig:net_LFM_ref}. Some findings are
consistent with ours. For example, disease 3 here is similar to
disease 1 from DFA. Both are related to LDL and TC. Likewise,
disease 2 is similar to disease 4 from DFA. But
 some findings remain difficult to interpret. 
Six  diseases are
associated with only one symptom. For instance, disease 10 is linked
to only MPV. However, high or low MPV alone is not of any clinical
importance. It  is only of concern if  other platelet-related
measurements are also out of their normal ranges. This synergy is well
captured by DFA (via disease 3).  \ech  

As already briefly commented in Section \ref{sec:intro}, graphical models may be also
employed for finding hidden structures of the symptoms. We run a
birth-death MCMC algorithm (available in the R package \texttt{BDgraph},
\citealt{mohammadi2015bdgraph}) for 50,000 iterations to learn a
Bayesian graphical model. 
A point estimate is shown in Figure \ref{fig:net_BD}. 
 Symptoms that form cliques of size  greater than 3 are
marked by circles. Some of the
findings are consistent with those by the DFA. As an illustration,
the clique of TC, triglycerides, LDL and HDL is very similar to
latent disease 1 in Figure \ref{fig:net_result}.
However, although graphical models can find
latent patterns that are not immediately obvious
in the correlation structure (Figure
\ref{fig:hm}a), like SLFM, it lacks inference on patient-disease
relationships. Moreover,  identifying cliques or other graph summaries
as disease  is an arbitrary choice. 

\section{Discussion}
\label{sec:disc}
We have developed a DFA model  for discovery of latent
diseases in EHR data.  
DFA can be equivalently viewed as  categorical matrix
factorization or as inference in 
edge-labeled random networks. 
In the EHR data analysis,  it is important to include available
diagnostic information and other prior knowledge,  which
greatly  facilitates identification of latent diseases. 
We found that the prevalence of damaged kidney function in our dataset is
 comparable with, but slightly  higher than the regional
average. We also
found 10 latent diseases that are related
to lipid disorder, thrombocytopenia, polycythemia, anemia, bacterial
and viral infections, allergy, and malnutrition.  
Finally, although the proposed DFA model is specifically designed for
disease mining, it could find broader applications in various 
areas such as education and psychology
\citep{chen2015statistical,chen2018bayesian}.

\bch As demonstrated in Section \ref{sec:cs}, inference under the
DFA model includes
inference on patient-disease relationships. Neither of the alternative
approaches, SLFM or graphical models, include this capability. Knowing
patient-disease relationships is crucial in prioritizing limited
medical resources.  The proposed inference for the DFA 
 is fully Bayesian.
Therefore, it outputs not only point estimates but also the
associated uncertainties through, e.g., the posterior probability of
symptom-disease relationships (reflected by the line width in Figure
\ref{fig:net_result}). SLFM is implemented using
an Expectation-Maximization (EM) algorithm. It can quickly get point estimates,
but does not include uncertainties.  

One of our preprocessing steps is discretizing the original
observations into binary or ternary variables. The main rationale of
discretization is two-fold: (1) the reference range of each test is
naturally incorporated; (2) it  reduces sensitivity to noise, outliers and distributional assumptions. The
downside of discretization, however, is a potential loss of
information, which can mitigated by increasing the number of categories
of the discretization. \ech  

Although EHR often involves  analysis of big datasets,  
scalability to large sample size is not the focus of this
paper. We focus on inference for data from a more narrowly restricted
subset equivalent to about a week worth of data.
We have shown that such data already allows meaningful inference on
latent diseases.
If desired and reasonable, the same approach could of course be used
for larger data sets, but would likely need to be 
combined with model extensions to allow for
changes in the latent structure, i.e., disease patterns, across 
different towns, times and other important factors.
Implementation needs computationally
efficient algorithms
for posterior inference such as consensus Monte Carlo. \bch In
Supplementary Material A, we  summarize   a small simulation study
assessing the performance of a consensus Monte Carlo algorithm for
 larger sample sizes. 
Computation time for a sample size of 15,000 is  under 5
minutes, and summaries of the estimation performance remain 
comparable to the MCMC algorithm
on small datasets. \ech

\bch Due to the conditional independence assumption of the sampling
model in \eqref{eqn:zij} and \eqref{eqn:cat}, given the model
parameters, the proposed DFA can be easily extended to inference on
datasets with missing values by simply ignoring factors in the
likelihood that involve missing values; similar approaches have been
studied extensively in the matrix completion and collaborative
filtering literature. One caveat is that it implicitly makes a
missing completely at random assumption.
The EHR dataset that
we considered here does not have massive missingness.
However, 
 more careful consideration of missingness is needed for application
to EHR data with possibly informative missingness. 
For instance, with laboratory test
results, a missing test item may suggest a normal result, if
one assumed that a test would only be ordered if based on other symptoms an
abnormal level were expected. \ech


\bibliography{DFA_ref}

\begin{thebibliography}{}

\bibitem[\protect\citename{Barrios {\em et~al.\ }\relax,
  }2013]{barrios&al:2013}
Barrios, Ernesto, Lijoi, Antonio, Nieto-Barajas, Luis~E., \& Prünster, Igor.
  2013.
\newblock Modeling with Normalized Random Measure Mixture Models.
\newblock {\em Statist. Sci.}, {\bf 28}(3), 313--334.

\bibitem[\protect\citename{Bhattacharya \& Dunson,
  }2011]{bhattacharya2011sparse}
Bhattacharya, Anirban, \& Dunson, David~B. 2011.
\newblock Sparse Bayesian infinite factor models.
\newblock {\em Biometrika}, {\bf 98}, 291--306.

\bibitem[\protect\citename{Broderick {\em et~al.\ }\relax,
  }2013]{broderick2013cluster}
Broderick, Tamara, Jordan, Michael~I, \& Pitman, Jim. 2013.
\newblock Cluster and feature modeling from combinatorial stochastic processes.
\newblock {\em Statistical Science}, {\bf 28}, 289--312.

\bibitem[\protect\citename{Campbell {\em et~al.\ }\relax,
  }2018]{campbell2018exchangeable}
Campbell, Trevor, Cai, Diana, \& Broderick, Tamara. 2018.
\newblock Exchangeable trait allocations.
\newblock {\em Electronic Journal of Statistics}, {\bf 12}(2), 2290--2322.

\bibitem[\protect\citename{Chang {\em et~al.\ }\relax, }2015]{RShiny}
Chang, Winston, Cheng, Joe, Allaire, JJ, Xie, Yihui, \& McPherson, Jonathan.
  2015.
\newblock {\em shiny: Web Application Framework for R}.
\newblock R package version 0.12.2.

\bibitem[\protect\citename{Chen {\em et~al.\ }\relax, }2018]{chen2018bayesian}
Chen, Yinghan, Culpepper, Steven~Andrew, Chen, Yuguo, \& Douglas, Jeffrey.
  2018.
\newblock Bayesian estimation of the DINA Q matrix.
\newblock {\em Psychometrika}, {\bf 83}(1), 89--108.

\bibitem[\protect\citename{Chen {\em et~al.\ }\relax,
  }2015]{chen2015statistical}
Chen, Yunxiao, Liu, Jingchen, Xu, Gongjun, \& Ying, Zhiliang. 2015.
\newblock Statistical analysis of Q-matrix based diagnostic classification
  models.
\newblock {\em Journal of the American Statistical Association}, {\bf
  110}(510), 850--866.

\bibitem[\protect\citename{Dahl, }2006]{dahl2006model}
Dahl, D.~B. 2006.
\newblock Model-Based Clustering for Expression Data via a {D}irichlet Process
  Mixture Model.
\newblock {\em In:} Vannucci, Marina, Do, Kim-Anh, \& M\"{u}ller, Peter (eds),
  {\em {B}ayesian Inference for Gene Expression and Proteomics}.
\newblock Cambridge University Press.

\bibitem[\protect\citename{De~Boer {\em et~al.\ }\relax, }2017]{de2017diabetes}
De~Boer, Ian~H, Bangalore, Sripal, Benetos, Athanase, Davis, Andrew~M, Michos,
  Erin~D, Muntner, Paul, Rossing, Peter, Zoungas, Sophia, \& Bakris, George.
  2017.
\newblock Diabetes and hypertension: a position statement by the American
  Diabetes Association.
\newblock {\em Diabetes Care}, {\bf 40}(9), 1273--1284.

\bibitem[\protect\citename{Dobra {\em et~al.\ }\relax,
  }2011]{dobra2011bayesian}
Dobra, Adrian, Lenkoski, Alex, \& Rodriguez, Abel. 2011.
\newblock Bayesian inference for general Gaussian graphical models with
  application to multivariate lattice data.
\newblock {\em Journal of the American Statistical Association}, {\bf
  106}(496), 1418--1433.

\bibitem[\protect\citename{Favaro \& Teh, }2013]{favaro&teh:2013}
Favaro, Stefano, \& Teh, Yee~Whye. 2013.
\newblock MCMC for Normalized Random Measure Mixture Models.
\newblock {\em Statist. Sci.}, {\bf 28}(3), 335--359.

\bibitem[\protect\citename{Goodfellow {\em et~al.\ }\relax,
  }2014]{goodfellow2014generative}
Goodfellow, Ian, Pouget-Abadie, Jean, Mirza, Mehdi, Xu, Bing, Warde-Farley,
  David, Ozair, Sherjil, Courville, Aaron, \& Bengio, Yoshua. 2014.
\newblock Generative Adversarial Nets.
\newblock {\em Pages  2672--2680 of:} {\em Advances in Neural Information
  Processing Systems}.

\bibitem[\protect\citename{Green \& Thomas, }2013]{green2011sampling}
Green, Peter~J., \& Thomas, Alun. 2013.
\newblock Sampling decomposable graphs using a Markov chain on junction trees.
\newblock {\em Biometrika}, {\bf 100}(1), 91--110.

\bibitem[\protect\citename{Griffiths \& Ghahramani,
  }2006]{ghahramani2006infinite}
Griffiths, Thomas~L., \& Ghahramani, Zoubin. 2006.
\newblock {Infinite latent feature models and the Indian buffet process}.
\newblock {\em Pages  475--482 of:} Weiss, Y., Sch\"olkopf, B., \& Platt, J.
  (eds), {\em Advances in Neural Information Processing Systems 18}.
\newblock MIT Press, Cambridge, MA.

\bibitem[\protect\citename{Griffiths \& Ghahramani, }2011]{griffiths2011indian}
Griffiths, Thomas~L, \& Ghahramani, Zoubin. 2011.
\newblock The Indian buffet process: An introduction and review.
\newblock {\em Journal of Machine Learning Research}, {\bf 12}(Apr),
  1185--1224.

\bibitem[\protect\citename{Guo, }2013]{guo2013bayesian}
Guo, Lei. 2013.
\newblock {\em Bayesian Biclustering on Discrete Data: Variable Selection
  Methods}.
\newblock Ph.D. thesis, Harvard University.

\bibitem[\protect\citename{Halpern {\em et~al.\ }\relax,
  }2016]{halpern2016electronic}
Halpern, Yoni, Horng, Steven, Choi, Youngduck, \& Sontag, David. 2016.
\newblock Electronic medical record phenotyping using the anchor and learn
  framework.
\newblock {\em Journal of the American Medical Informatics Association}, {\bf
  23}(4), 731--740.

\bibitem[\protect\citename{Hartigan, }1972]{hartigan1972direct}
Hartigan, John~A. 1972.
\newblock Direct clustering of a data matrix.
\newblock {\em Journal of the American Statistical Association}, {\bf 67}(337),
  123--129.

\bibitem[\protect\citename{Henderson {\em et~al.\ }\relax,
  }2017]{henderson2017granite}
Henderson, Jette, Ho, Joyce~C, Kho, Abel~N, Denny, Joshua~C, Malin, Bradley~A,
  Sun, Jimeng, \& Ghosh, Joydeep. 2017.
\newblock Granite: Diversified, sparse tensor factorization for electronic
  health record-based phenotyping.
\newblock {\em Pages  214--223 of:} {\em Healthcare Informatics (ICHI), 2017
  IEEE International Conference on}.
\newblock IEEE.

\bibitem[\protect\citename{Lau \& Green, }2007]{lau2007bayesian}
Lau, John~W, \& Green, Peter~J. 2007.
\newblock Bayesian model-based clustering procedures.
\newblock {\em Journal of Computational and Graphical Statistics}, {\bf 16}(3),
  526--558.

\bibitem[\protect\citename{Lauritzen, }1996]{lauritzen1996graphical}
Lauritzen, Steffen~L. 1996.
\newblock {\em Graphical models}.
\newblock  Vol. 17.
\newblock Oxford: Clarendon Press.

\bibitem[\protect\citename{Lee {\em et~al.\ }\relax, }2015]{lee2015}
Lee, Juhee, Müller, Peter, Gulukota, Kamalakar, \& Ji, Yuan. 2015.
\newblock A Bayesian feature allocation model for tumor heterogeneity.
\newblock {\em Ann. Appl. Stat.}, {\bf 9}(2), 621--639.

\bibitem[\protect\citename{Li, }2005]{li2005general}
Li, Tao. 2005.
\newblock A general model for clustering binary data.
\newblock {\em Pages  188--197 of:} {\em Proceedings of the eleventh ACM SIGKDD
  international conference on Knowledge discovery in data mining}.
\newblock ACM.

\bibitem[\protect\citename{Meeds {\em et~al.\ }\relax,
  }2007]{meeds2007modeling}
Meeds, Edward, Ghahramani, Zoubin, Neal, Radford~M, \& Roweis, Sam~T. 2007.
\newblock Modeling dyadic data with binary latent factors.
\newblock {\em Pages  977--984 of:} {\em Advances in Neural Information
  Processing Systems}.

\bibitem[\protect\citename{Miettinen {\em et~al.\ }\relax,
  }2008]{miettinen2008discrete}
Miettinen, Pauli, Mielik{\"a}inen, Taneli, Gionis, Aristides, Das, Gautam, \&
  Mannila, Heikki. 2008.
\newblock The discrete basis problem.
\newblock {\em IEEE Transactions on Knowledge and Data Engineering}, {\bf
  20}(10), 1348--1362.

\bibitem[\protect\citename{Miller \& Harrison, }2018]{Miller2017}
Miller, Jeffrey~W., \& Harrison, Matthew~T. 2018.
\newblock Mixture Models With a Prior on the Number of Components.
\newblock {\em Journal of the American Statistical Association}, {\bf
  113}(521), 340--356.

\bibitem[\protect\citename{Mohammadi \& Wit, }2015]{mohammadi2015bdgraph}
Mohammadi, Abdolreza, \& Wit, Ernst~C. 2015.
\newblock BDgraph: an R package for Bayesian structure learning in graphical
  models.
\newblock {\em arXiv preprint arXiv:1501.05108}.

\bibitem[\protect\citename{Ni {\em et~al.\ }\relax, }2018]{ni2018scalable}
Ni, Yang, M{\"u}ller, Peter, Diesendruck, Maurice, Williamson, Sinead, Zhu,
  Yitan, \& Ji, Yuan. 2018.
\newblock Scalable Bayesian Nonparametric Clustering and Classification.
\newblock {\em arXiv preprint arXiv:1806.02670}.

\bibitem[\protect\citename{Pontes {\em et~al.\ }\relax,
  }2015]{pontes2015biclustering}
Pontes, Beatriz, Gir{\'a}ldez, Ra{\'u}l, \& Aguilar-Ruiz, Jes{\'u}s~S. 2015.
\newblock Biclustering on expression data: A review.
\newblock {\em Journal of Biomedical Informatics}, {\bf 57}, 163--180.

\bibitem[\protect\citename{Richardson \& Green, }1997]{richardson1997bayesian}
Richardson, Sylvia, \& Green, Peter~J. 1997.
\newblock On Bayesian analysis of mixtures with an unknown number of components
  (with discussion).
\newblock {\em Journal of the Royal Statistical Society: series B (statistical
  methodology)}, {\bf 59}(4), 731--792.

\bibitem[\protect\citename{Ro{\v{c}}kov{\'a} \& George,
  }2016]{rovckova2016fast}
Ro{\v{c}}kov{\'a}, Veronika, \& George, Edward~I. 2016.
\newblock Fast Bayesian factor analysis via automatic rotations to sparsity.
\newblock {\em Journal of the American Statistical Association}, {\bf
  111}(516), 1608--1622.

\bibitem[\protect\citename{Ross {\em et~al.\ }\relax, }2014]{ross2014big}
Ross, MK, Wei, Wei, \& Ohno-Machado, L. 2014.
\newblock "Big data" and the electronic health record.
\newblock {\em Yearbook of Medical Informatics}, {\bf 9}(1), 97.

\bibitem[\protect\citename{Rukat {\em et~al.\ }\relax,
  }2017]{rukat2017bayesian}
Rukat, Tammo, Holmes, Chris~C, Titsias, Michalis~K, \& Yau, Christopher. 2017.
\newblock Bayesian boolean matrix factorisation.
\newblock {\em arXiv preprint arXiv:1702.06166}.

\bibitem[\protect\citename{Scott \& Berger, }2010]{scott2010bayes}
Scott, James~G, \& Berger, James~O. 2010.
\newblock {Bayes and empirical-Bayes multiplicity adjustment in the
  variable-selection problem}.
\newblock {\em The Annals of Statistics}, {\bf 38}, 2587--2619.

\bibitem[\protect\citename{Uitert {\em et~al.\ }\relax,
  }2008]{uitert2008biclustering}
Uitert, Miranda~van, Meuleman, Wouter, \& Wessels, Lodewyk. 2008.
\newblock Biclustering sparse binary genomic data.
\newblock {\em Journal of Computational Biology}, {\bf 15}(10), 1329--1345.

\bibitem[\protect\citename{Wood {\em et~al.\ }\relax, }2006]{wood2006npb}
Wood, F., Griffiths, T.~L., \& Ghahramani, Z. 2006.
\newblock {A non-parametric Bayesian method for inferring hidden causes}.
\newblock {\em In:} {\em Proceedings of the Conference on Uncertainty in
  Artificial Intelligence},  vol. 22.

\bibitem[\protect\citename{Xu {\em et~al.\ }\relax, }2013]{xu2013nonparametric}
Xu, Yanxun, Lee, Juhee, Yuan, Yuan, Mitra, Riten, Liang, Shoudan, M{\"u}ller,
  Peter, \& Ji, Yuan. 2013.
\newblock Nonparametric Bayesian bi-clustering for next generation sequencing
  count data.
\newblock {\em Bayesian Analysis}, {\bf 8}(4), 759--780.

\bibitem[\protect\citename{Zhang {\em et~al.\ }\relax,
  }2017]{zhang2017hypertension}
Zhang, Fu-Liang, Guo, Zhen-Ni, Xing, Ying-Qi, Wu, Yan-Hua, Liu, Hao-Yuan, \&
  Yang, Yi. 2017.
\newblock Hypertension prevalence, awareness, treatment, and control in
  northeast China: a population-based cross-sectional survey.
\newblock {\em Journal of Human Hypertension}, {\bf 32}(1), 54--65.

\bibitem[\protect\citename{Zhang {\em et~al.\ }\relax,
  }2012]{zhang2012prevalence}
Zhang, Luxia, Wang, Fang, Wang, Li, Wang, Wenke, Liu, Bicheng, Liu, Jian, Chen,
  Menghua, He, Qiang, Liao, Yunhua, Yu, Xueqing, {\em et~al.\ }\relax. 2012.
\newblock Prevalence of chronic kidney disease in China: a cross-sectional
  survey.
\newblock {\em The Lancet}, {\bf 379}(9818), 815--822.

\bibitem[\protect\citename{Zhang {\em et~al.\ }\relax, }2007]{zhang2007binary}
Zhang, Zhongyuan, Li, Tao, Ding, Chris, \& Zhang, Xiangsun. 2007.
\newblock Binary matrix factorization with applications.
\newblock {\em Pages  391--400 of:} {\em Data Mining, 2007. ICDM 2007. Seventh
  IEEE International Conference on}.
\newblock IEEE.

\end{thebibliography}
\end{document}